\documentclass[preprint,aps,tightenlines,nofootinbib]{revtex4}
\usepackage{graphicx,amsfonts,amssymb,amsmath}

\renewcommand{\bar}{\overline}

\newcommand{\nc}{\newcommand}
\nc{\grad}{\nabla}  
\nc{\tr}{\mathop{\rm Tr}\,}
\nc{\half}{{1\over 2}}
\nc{\third}{{1\over 3}}
\nc{\be}{\begin{equation}}
\nc{\ee}{\end{equation}}
\nc{\bea}{\begin{eqnarray}}
\nc{\eea}{\end{eqnarray}}

\nc{\dint}[2]{\int\limits_{#1}^{#2}}
\nc{\D}{\displaystyle}
\nc{\PDT}[1]{\frac{\partial #1}{\partial t}}
\nc{\tw}{\tilde{w}}
\nc{\tg}{\tilde{g}}
\nc{\newcaption}[1]{\centerline{\parbox{5.6in}{\caption{#1}}}}
\def\href#1#2{#2} 

\def\beq{\begin{eqnarray}}   \def\eeq{\end{eqnarray}}

\def\lsim{\mathrel{\rlap{\lower3pt\hbox{\hskip0pt$\sim$}}
    \raise1pt\hbox{$<$}}}         %less than or approx. symbol
\def\gsim{\mathrel{\rlap{\lower4pt\hbox{\hskip1pt$\sim$}}
    \raise1pt\hbox{$>$}}}         %greater than or approx. symbol

\def\Z{{\mathbb{Z}}}
\def\C{{\mathbb{C}}}
\def\R{{\mathbb{R}}}
\def\Id{\hbox{1\kern-.23em{\rm l}}}

\def\lsim{\mathrel{\rlap{\lower3pt\hbox{\hskip0pt$\sim$}}
    \raise1pt\hbox{$<$}}}         
\def\gsim{\mathrel{\rlap{\lower4pt\hbox{\hskip1pt$\sim$}}
    \raise1pt\hbox{$>$}}}         
\nc{\al}{\alpha}
\nc{\ga}{\gamma}
\nc{\de}{\delta}
\nc{\ep}{\epsilon}
\nc{\ze}{\zeta}
\nc{\et}{\eta}
\renewcommand{\th}{\theta}
\nc{\Th}{\Theta}
\nc{\ka}{\kappa}
\nc{\la}{\lambda}
\nc{\rh}{\rho}
\nc{\si}{\sigma}
\nc{\ta}{\tau}
\nc{\up}{\upsilon}
\nc{\ph}{\phi}
\nc{\ch}{\chi}
\nc{\ps}{\psi}
\nc{\om}{\omega}
\nc{\Ga}{\Gamma}
\nc{\De}{\Delta}
\nc{\La}{\Lambda}
\nc{\Si}{\Sigma}
\nc{\Up}{\Upsilon}
\nc{\Ph}{\Phi}
\nc{\Ps}{\Psi}
\nc{\Om}{\Omega}
\nc{\ptl}{\partial}
\nc{\del}{\nabla}
\nc{\ov}{\overline}
\nc{\gsl}{\!\not}
\nc{\bi}[1]{\bibitem{#1}}
\nc{\fr}[2]{\frac{#1}{#2}}
\nc{\dsl}{\partial\!\!\!\!\!\!\not\,\,}
\nc{\gm}{\mbox{$\gamma_{\mu}$}}
\nc{\gn}{\mbox{$\gamma_{\nu}$}}
\nc{\Le}{\mbox{$\fr{1+\gamma_5}{2}$}}
\nc{\Ri}{\mbox{$\fr{1-\gamma_5}{2}$}}
\nc{\GD}{\mbox{$\tilde{G}$}}
\nc{\gf}{\mbox{$\gamma_{5}$}}
\nc{\Ima}{\mbox{Im}}
\nc{\Rea}{\mbox{Re}}

\nc{\av}{\langle \ph\rangle}
\nc{\ntwo}{${\cal N}\!\!=\!2\;$}
\nc{\none}{${\cal N}\!\!=\!1\;$}
\nc{\nfour}{${\cal N}\!\!=\!4\;$}
\nc{\cp}{\C{\rm P}}

\def \bi{\bibitem}
\nc{\rf}[1]{(\ref{#1})}
\def \del{\partial}

\begin{document}
%\draft
\begin{flushright}
{ CERN-PH-TH/2004-092 \\
FTPI-MINN-04/17\\
UMN-TH-2306/04}
\end{flushright}

\setcounter{page}{1}

\vspace*{1in}
\title{Enhanced Worldvolume Supersymmetry and \\Intersecting Domain Walls in 
\boldmath{\none} SQCD \\ $\;\;$ \\}

\author{Adam Ritz,$^{a}$ Mikhail Shifman,$^{b}$ and
  Arkady Vainshtein$^{\,b}$}
\affiliation{
$^a$Theory Division, Department of Physics, CERN,
Geneva 23, CH-1211, Switzerland\\
$^b$William I. Fine Theoretical Physics Institute, University of Minnesota,
116 Church St SE,\\ Minneapolis, MN 55455, USA \\ $\;\;$ \\
}

\begin{abstract} $\;\;$ \\
  
We study the worldvolume dynamics of BPS domain walls in 
\none SQCD with $N_f=N$ flavors, and exhibit an enhancement of supersymmetry for the
reduced moduli space associated with broken flavor symmetries. We provide an explicit 
construction of the worldvolume superalgebra which corresponds to an 
\ntwo K\"ahler sigma model in 2+1D deformed by a potential, given by the 
norm squared of a U(1) Killing vector, resulting from
the flavor symmetries broken by unequal quark masses.
This framework leads to a worldvolume description of novel
two-wall junction configurations, which are 1/4-BPS objects, but
nonetheless preserve two supercharges when viewed as kinks on
the wall worldvolume. 

\end{abstract}

\maketitle
\thispagestyle{empty}
\newpage

%%%%%%%%%%%%%%%%%%%%%%%%%%%

\section{Introduction}

One of the more profound features of supersymmetric field theories 
is that solitonic field configurations are often endowed with a special status,
namely they are annihilated by a certain number of supercharges and thus
lie in shortened, or BPS, representations \cite{A4}. This feature has far-reaching consequences
due to the ensuing non-renormalization theorems which affect the mass  (or tension)
and spectrum of these solitons, some of which may play an important role in the dynamics.
In general, soliton configurations exhibit a moduli space of solutions,
and much insight can be gleaned from a study of the low energy collective
coordinate dynamics on this space and its induced metric \cite{manton1}. This 
is particularly true in cases where the moduli space is 
nontrivial in the sense that it includes 
components beyond that associated with the broken translation generators; the
latter component is always present on the grounds that a soliton is a localized 
configuration. 

In the supersymmetric context, the moduli space ${\cal M}$ {\it locally}
admits the general decomposition,
\be
 {\cal M} \simeq {\cal M}_{\rm SUSY} \times \widetilde{\cal M}\ ,
\ee
where ${\cal M}_{\rm SUSY}$ refers to the sector associated with bosonic 
generators in the supersymmetry (SUSY) algebra which are broken by the soliton, and
in flat space always includes a translational component $\R^d \subset {\cal M}_{\rm SUSY}$,
where $d$ is the codimension. The realization of supersymmetry in this sector,
associated with the unbroken generators, is then fixed by the kinematics of 
the bulk superalgebra.

In contrast, $\widetilde{\cal M}$ -- the `reduced moduli space' -- is not 
directly associated with broken generators in the superalgebra. This has
the important consequence that in certain cases the realization of 
worldvolume supersymmetry is less constrained by the bulk kinematics.
In particular, we will argue here that there are
situations in which the number of supercharges which act trivially on the reduced
moduli space of a BPS soliton can be {\it larger} than one 
would infer directly from the preserved fraction of bulk supersymmetry. 
The origin of this supersymmetry enhancement is that not all
of the supercharges which are realized on the worldvolume of the soliton 
lift to supercharges in the full theory. The additional {\it supernumerary} 
supercharges arise due to special geometric features of the reduced moduli 
space, e.g. a K\"ahler or hyperK\"ahler structure, which are 
not present within the full theory. 

The primary aim of this paper is to illustrate how this novel feature 
plays an important role in the dynamics of 1/2-BPS domain walls in
\none SQCD. In particular, we will focus on the theory with gauge group 
SU($N$) accompanied by $N_f=N$ fundamental flavors with masses which are small
relative to the dynamical scale, $\La_N$, of the theory. This theory 
has a low energy description
on the Higgs branch, in terms of meson and baryon chiral superfield moduli, 
where it reduces to a massive perturbation of a K\"ahler sigma model on the
manifold determined by the quantum constraint \cite{Seiberg},
\be
 {\rm det}M - B\tilde{B} = \La_N^{2N}\ .
\ee
The massive theory possesses $N$ quantum vacua which, with a 
  hierarchical structure for the quark mass matrix, are in the
  weak coupling regime.
On decoupling $N$ flavors, these $N$ vacua tend smoothly to the $N$ quantum vacua
of pure \none SYM \cite{Witten1,Veneziano,ADS,nsvz,SVone}.

\begin{figure}[h]
\includegraphics[width=5cm]{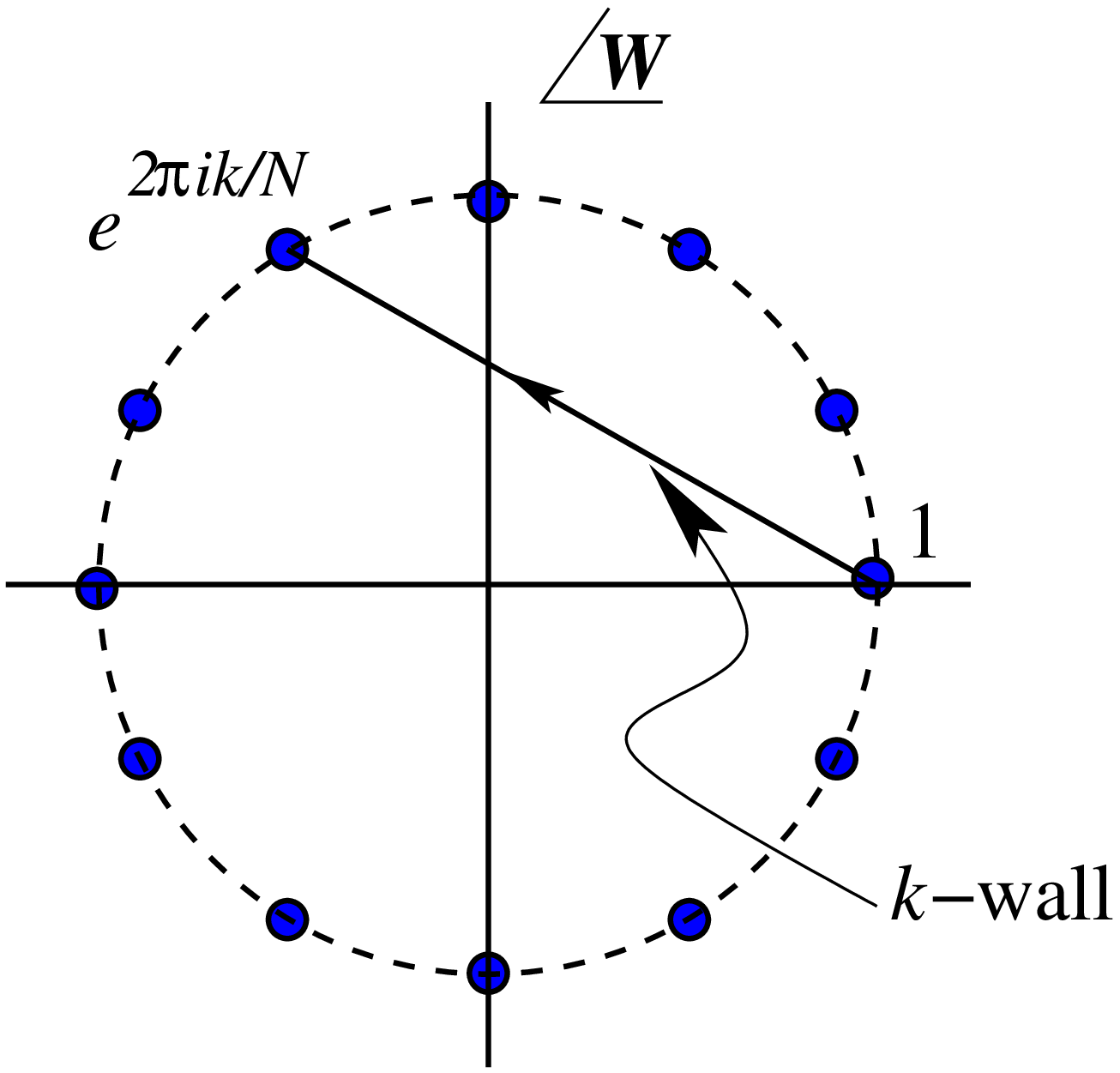}
\vspace*{0.3cm}
 \newcaption{\footnotesize A schematic representation of the $N$ vacua, and a 
$k$-wall, for \none SQCD with $N$ flavors.}
\end{figure}

The $N$ distinct vacua of this theory allow for domain wall solutions which interpolate
between them. The corresponding central charge is present in the superalgebra
\cite{DSone} and such solitons are 1/2-BPS saturated. In previous work \cite{rsv}, we
studied the BPS wall spectrum in this theory, following earlier work on BPS walls
in other variants of \none SQCD \cite{DSone,Stwo,CS,walls1,kaplunovsky,wallsN,CM,CHMM}.
The vacuum structure is illustrated in the plane of the superpotential in Fig.~1, 
which also provides a graphical definition of a $k$-wall, namely a BPS wall which
interpolates between vacua differing in phase by $2\pi k/N$. In \cite{rsv} we argued,
as reviewed below, that $k$-walls exhibit a nontrivial classical reduced moduli space
$\widetilde{\cal M}_k$ due to localized Goldstone modes associated with the
flavor symmetries which are broken by the wall solution. The 
corresponding coset is a complex Grassmannian \cite{rsv},
\be
 \widetilde{\cal M}_k = G(k,N) \equiv 
 \frac{{\rm U}(N)}{{\rm U}(k) \times {\rm U}(N-k)} \ . 
\ee
One can then formally deduce that the multiplicity of $k$-walls, $\nu_k$, is given by
the worldvolume Witten index for this Grassmannian sigma model, which depends only
on the topology of the space, and is given by the Euler characteristic,
\be
 \nu_k ~=~ \ch(G(k,N)) ~=~ \frac{N!}{k!(N-k)!}\ . \label{degen1}
\ee
This was the primary result of \cite{rsv}, which interestingly was
consistent with an alternative string-theoretic picture of BPS walls in pure 
\none SYM \cite{av}.

In the present paper, we wish to study the worldvolume dynamics in more detail,
and resolve some of the puzzles which arise from a closer inspection of the above result.
One of these is the statement that the reduced moduli space is a K\"ahler manifold.
Since the worldvolume theory lives in 2+1D, the dual constraints of 
(i) a K\"ahler target space,
and (ii) Lorentz invariance, imply that the low energy dynamics must preserve \ntwo 
supersymmetry, namely four supercharges! Since only two bulk supercharges act trivially
on the soliton solution, this conclusion clearly requires some justification. 
A seemingly related paradox was in fact noted some time ago in considering the 
K\"ahler moduli space of lumps in K\"ahler sigma models \cite{ruback}. However, in 
the latter case, the problem dissipates once one realizes that Lorentz invariance
places no constraint and one can consistently realize just two supercharges
in terms of one-component fermions \cite{gauntlett92}. The situation here 
allows for no such resolution and, as alluded to above, in this 
case there is indeed an enhancement of supersymmetry, at least at the two-derivative level. 
This enhancement does not of course apply to the (decoupled) translational sector, but only
to the reduced moduli space. We will provide an explicit example of how this can 
occur, and then apply it to $k$-walls in SQCD and more specifically to
the simplest case of 1-walls in the case of an SU(2) gauge group. We note that
the mechanism appears likely to apply more widely for other solitons in \none theories.

A second issue that we aim to resolve is to understand what happens to the
flavor moduli coordinatizing $\widetilde{\cal M}_k$ when we explicitly break
some of the flavor symmetries by putting the quark mass matrix in a hierarchical
form. In such a maximally asymmetric regime, the wall no longer breaks any additional 
global symmetries and one anticipates that the moduli space should be lifted.
We will provide evidence that this is indeed the case. In particular, by
considering the realization of the worldvolume supercharges for the 
SU(2) $N_f=2$ theory, we show that, for a linear order perturbation in the mass matrix,
the effect is to introduce a potential on the moduli space which geometrically
is the norm squared of a U(1) Killing vector. Such a `real mass' 
deformation in 2+1D is known to
be consistent with \ntwo SUSY \cite{agf}.
Moreover, one important consistency check is that the result one obtains via 
this linear deformation is in fact perfectly
compatible with the opposite limit in which the second flavor is integrated out.

The third and final aim of this work is to explore the realization of other bulk solitons
within the worldvolume theory of domain walls. The example we focus on corresponds to a novel
class of two-wall 1/4-BPS junctions which are possible by virtue of the degeneracy
(\ref{degen1}) of BPS walls interpolating between the same two vacua. We will
provide evidence that these configurations can be identified with 1/2-BPS kinks in 
the worldvolume Grassmannian sigma model. These configurations thus preserve two worldvolume
supercharges only one of which can be identified with the unbroken bulk supercharge. 
As evidence for this identification,
we will verify for the SU(2) case that there is a direct match for 
the tension between the bulk result obtained in the hierarchical mass regime, and
the appropriate limit of the kink tension for the massive sigma model.

This paper is organized as follows. In the next section we consider
the moduli space of BPS solitons, and discuss in some generality the 
worldvolume realization of supersymmetry. We argue that the
reduced moduli space may in certain cases exhibit supersymmetry
enhancement and present a simple sigma model where this arises
for the worldline dynamics of BPS kinks.
We then turn to the specific case of BPS walls in $N_f=N$ flavor SQCD
in Sect.~3, recalling the structure of the wall moduli space \cite{rsv}
and then describing the worldvolume realization of supersymmetry,
which on the reduced moduli space is enhanced to \ntwo. We describe the 
structure explicitly for the SU(2) case with unequal quark masses, as is 
required to remain at weak coupling.
This viewpoint is applied in Sect.~4 to consider novel 1/4-BPS two-wall 
junctions from the viewpoint of the wall worldvolume. We finish
with some concluding remarks on other worldvolume solitons, including lumps, in
Sect.~5. In an appendix, we review the structure of tensorial central charges
in $D=2,3$ and 4, noting a subtlety with vectorial string charges.

\section{Supersymmetry and worldvolume moduli}

In this section we will discuss some aspects of the matching between
bosonic and fermionic moduli for BPS solitons. We distinguish the 
translational sector, which is essentially fixed on kinematic
grounds, from the remainder of the moduli -- the reduced moduli space --
which we argue, by way of an explicit example, can in 
certain situations exhibit an `enhancement' of supersymmetry, 
in the sense that the associated dynamics preserves more supercharges than one
would infer from the bulk superalgebra. The additional supercharges act only on the
reduced moduli space and are not present in the bulk theory.

\subsection{Counting moduli and the translational sector}

We begin with a simple physical perspective on the matching between bosonic and 
fermionic moduli of BPS solitons. Recall that on general grounds the bosonic moduli 
space for a
configuration of solitons in flat space locally admits the decomposition
\be
 {\cal M} \simeq {\cal M}_{\rm SUSY} \times \widetilde{\cal M} \ , \label{decomp}
\ee
where ${\cal M}_{\rm SUSY}$ is the sector associated with broken 
(bosonic) symmetry generators in the superalgebra. The second factor in (\ref{decomp}), 
$\widetilde{\cal M}$, encodes any other modes associated with broken global 
symmetries, e.g. relative translations or, as will be more relevant here, flavor symmetries.

Consider a bosonic soliton configuration ${\cal S}(x)$ in $D$-dimensional Minkowski space
which has finite mass (or tension) -- large relative to the scales of the underlying
theory -- and is localised in $d\leq D-1$ spatial dimensions. 
Within a Lorentz invariant field theory, it is clear that this configuration 
possesses $d$ localized bosonic zero modes as it spontaneously breaks  
translational invariance. It follows that the minimal content of the 
moduli space takes the form
\be
 {\cal M}_{\rm SUSY}^{\rm min} = \R^d\ .
\ee
For solitons within theories of extended supersymmetry, 
${\cal M}_{\rm SUSY}$ may acquire additional bosonic dimensions, due to
the enforced K\"ahler or hyperK\"ahler structure.

We would now like to argue that there are {\it at least} $d$ fermionic zero modes
of the soliton configuration if the bulk theory possesses linearly realized
supersymmetry. More precisely, we will consider a soliton in a globally\footnote{Exceptions to 
this correspondence are known within supergravity \cite{sugra}, where the condition that 
Killing spinors be normalizable at infinity becomes nontrivial.} supersymmetric
field theory. Furthermore, to simplify the discussion, we will assume a real representation for 
the superalgebra.\footnote{This includes the cases $D=2,3$ and 4
which we will focus on here, but the argument should generalize appropriately
to dimensions without Majorana spinors.} Since two SUSY variations
commute to a translation, $[\de_1,\de_2]S(x) \propto \ga^\mu \ptl_\mu\, S(x)$, it
follows that if there are $d$ broken translational generators, there are at least
$d$ broken supersymmetry generators, and thus $d$ fermionic zero modes. In 
practice, the number may of course be larger on account of Lorentz invariance or
extended SUSY. 

While this matching is essentially enforced by the representation theory of the worldvolume
superalgebra in many examples, it is interesting that this simple argument 
also applies even if the worldvolume of the soliton is 0+1-dimensional, for which 
supersymmetric quantum mechanics in principle imposes no fixed relation between the 
number of bosonic and fermionic degrees of freedom. Moreover, we did not need to 
assume the existence of a well-defined fermion parity $(-1)^F$ on the states, and 
thus representations could exist which are not Bose-Fermi paired. 
As simple illustrations of the minimal one-to-one matching
consider first a putative BPS vortex in a theory with minimal
\none SUSY (i.e. two supercharges) in 2+1D. Such a configuration would require a 
worldvolume description with two bosonic zero modes, but only one fermionic mode.
This is not permitted by the argument above, and indeed no
configurations of this type are known. Vortices always exhibit
at least two fermionic zero modes and are thus BPS only in \ntwo theories
in 2+1D. As a second example, consider an SU(2) Yang-Mills instanton. Within
the \none superconformal algebra they possess eight bosonic, but
only four fermionic, zero modes, and are indeed BPS. However,
to exist as solitons we must lift them to 4+1D where the 
minimal superconformal algebra possesses sixteen generators 
and BPS instantons then exhibit eight fermionic zero modes restoring 
the minimal one-to-one matching.

Proceeding further, one notes that a one-to-one matching between
bosonic and fermionic modes (in practice a two-to-one matching of 
phase space variables) is possible only in the
absence of nontrivial constraints from Lorentz invariance,
namely when the worldvolume is 0+1 or 1+1-dimensional. 
These cases still cover the majority of solitons present within
theories in 3+1D, and this minimal matching is known to occur 
in many cases. The first example where Lorentz invariance
does impose a constraint arises for BPS walls in 3+1D, where
the mode matching must be one-to-two. It is this case that will
be of interest here.

The discussion above focused on the translational or, more generally, the
`super-Goldstone' sector of the moduli space ${\cal M}_{\rm SUSY}$. 
The constraints imposed by Lorentz invariance on $\widetilde{\cal M}$ are the same, 
but in general the realization of supersymmetry may be somewhat different. 
This is the issue to which we turn next.

\subsection{Supersymmetry enhancement for \boldmath$\widetilde{\cal M}$: 
the \none $S^3$ sigma model}

The realization of SUSY in the translational sector of the moduli space
is highly constrained by kinematics. In contrast, the reduced moduli
space may contain moduli which are unrelated to translational
zero modes and the structure of the superalgebra. Of course,
this is not necessarily the case if we consider a multi-soliton configuration
where $\widetilde{\cal M}$ will include moduli corresponding to relative translations, 
but we have in mind a situation where $\widetilde{\cal M}$
is instead associated with other broken global symmetries.
In this case, we will argue that the reduced moduli space may exhibit
an apparent `enhancement' of supersymmetry at the two-derivative
level relative to the full dynamics on ${\cal M}$. 

To motivate why supersymmetry enhancement for the low energy dynamics on
$\widetilde{\cal M}$ can be rather natural, we will first present an explicit example. 
Consider an \none sigma model in 1+1D with target space $S^3$ 
accompanied by its round metric \cite{lsv},
for which we introduce spherical polar coordinates $\ph^a=\{\th,\xi,\ph\}$,
\be
 ds^2 = r \left[ d\th^2 + \sin^2\th \left( d\xi^2 + \sin^2\xi d\ph^2\right)
   \right].
\ee
We also turn on a (real) superpotential,
\be
 {\cal W}(\ph) = m \cos\th\ ,
\ee
which depends on only one of the angular coordinates parametrizing
the $S^3$. The theory then has two vacua at $\th=0,\pi$.

Classical BPS kinks exist which interpolate between the two vacua,
having mass
\be
 M_{\rm sol} = {\cal Z} = 2m\ ,
\ee
and satisfying the Bogomol'nyi equation,
\be
 \ptl_z \ph^a = g^{ab} \ptl_b {\cal W}(\ph)\ . \label{bogeqn}
\ee
The solutions have the simple sine-Gordon form
\be
 \th_{\rm sol}(z) = 2\arctan\left[\exp\left(-\frac{m}{r}(z-z_0)\right)\right], \;\;\;\;\;\;
   \xi_{\rm sol}=\xi_0,\;\;\;\;\;\;  \ph_{\rm sol}=\ph_0\ , \label{sgsol}
\ee
exhibiting three bosonic moduli $\{z_0,\xi_0,\ph_0\}$. 

These bosonic moduli are Goldstone modes for the symmetries 
broken by the wall: $z_0$ is associated with the breaking of
translation invariance; $\xi_0$ and $\ph_0$ arise
from the SO(3) global symmetry of the target space which is
preserved in the vacua but broken to SO(2) by the kink solution.
We thus anticipate that $\xi_0$ and $\ph_0$ coordinatize the
coset SO(3)/SO(2) $\simeq S^2$. This may be verified by 
inserting the solution into the bosonic Lagrangian and 
computing the induced metric for the bosonic zero modes \cite{manton1}, on allowing
for weak time-dependence,
\be
\label{coord}
 ds^2_{\cal M} =2 m\, dz_0^2+ h_{ij} dx^i dx^j = 
  2 m\, dz_0^2 + \frac{2r^2}{m} \left[ d\xi_0^2 + \sin^2\xi 
   d \ph_0^2\right]\ , \qquad i,j=1,2\ ,
\ee
where $h_{ij}$ is the metric of the reduced moduli space $\widetilde{\cal M}$.
The bosonic moduli space is thus
\be
 {\cal M} = \R \times \widetilde{\cal M} = \R \times S^2\ ,
\ee
with the natural metric on each factor.

Let us now consider the fermionic sector. The $S^3$ coordinates
$\ph^a$ are partnered under \none SUSY by a set of two-component
Majorana spinors, $\ps^a_{\al}$, $\al=1,2$. For each bosonic
zero mode $x^i$, one finds a corresponding (one-component) 
fermionic partner $\et^i$ in the lower component of $\ps^a_{\al}$,
\be
 \ps^a_{\rm sol} = \et^i \frac{\ptl \ph^a_{\rm sol}}{\ptl x^i} 
   \left(\begin{array}{c} 0 \\ 1 \end{array}\right) + \mbox{non-zero modes}\ .
\ee
Only one of these modes is guarunteed to exist by virtue of the fact that the solution 
is classically 1/2-BPS and thus breaks one of the two supercharges.
The broken supercharge is realized as 
\be
 Q_1 = 2{\cal Z} \et^z\ ,
\ee
in terms of this `goldstino' mode. Here $\et^z$ is the superpartner of $z_0$.

We now come to a rather surprising feature of this system.
The reduced moduli space $\widetilde{\cal M}$ is a K\"ahler manifold
and, since the bosonic and fermionic zero modes are paired, exhibits
\ntwo supersymmetry. One of these supercharges is $Q_2$, the unbroken
charge present in the bulk theory, while the second which we will
call $\widetilde{Q}_2$ exists only due to the complex structure
$J$ associated with $\widetilde{\cal M}$.  In the coordinate system (\ref{coord}),
\be
J=\left(\begin{array}{cc}0&1\\-\!1\;& 0\end{array}\right)\ . \label{compstruc}
\ee
We can represent the
supercharges as\,\footnote{Implying that the algebra is restricted to the rest frame, 
we take $\dot z_{0}=0$.}
\bea
 {\cal Q}^I \equiv \left\{ \begin{array}{l} 
                               Q_2 = h_{ij} \dot{x}^i \et^j \\[1mm]
                  \widetilde{Q}_2 = h_{ij} J^j_k \dot{x}^i \et^k
                           \end{array} \right. 
\eea
and, noting that $\{\et^i,\et^j\}=h^{ij}$, one can verify that
they satisfy the algebra of \ntwo SQM,
\be
 \{{\cal Q}^I, {\cal Q}^J\} = {\cal H}_{\rm SQM} \de^{IJ}\ ,
\ee
where ${\cal H}_{\rm SQM} = (M-{\cal Z})$ is the worldline
Hamiltonian. Introducing the complex coordinate 
\be
w={\rm e}^{i\phi_{0}}\tan\frac{\xi_0}{2} \label{ccoord}
\ee
on $\widetilde{\cal M}$, and its fermionic partner
\be
 \ps = \frac{1}{2} \,\sec^2\frac{\xi_0}{2} e^{i\ph_0} \Big( \et^{\xi} 
 + i \sin\xi_0 \et^{\ph}\Big),
\ee
we can  rewrite the algebra in the form
\be
 \{{\cal Q},{\cal Q}^*\} = {\cal H}_{\rm SQM}\ , \;\;\;\;\;\;\ 
 ({\cal Q})^2 = ({\cal Q}^*)^2 = 0\ ,
\ee
where
\bea
 {\cal Q} &=& \frac{1}{2}({\cal Q}^1 + i {\cal Q}^2) = h_{w\bar{w}} \dot{\bar{w}}\ps\ , 
 \nonumber \\[1mm]
{\cal Q}^*\! \!&=& \frac{1}{2}({\cal Q}^1 - i {\cal Q}^2) = h_{w\bar{w}} \dot{w}\bar\ps\ .
\eea

At this point we should emphasize that the arguments for enhanced supersymmetry 
presented above refer to the low energy or two-derivative sector of the worldvolume theory. 
Since supersymmetry does not enforce this enhancement, nor indeed the K\"ahler structure of the
reduced moduli space, it seems inevitable that higher derivative terms
on the worldvolume will not respect \ntwo supersymmetry. We will not attempt to
verify this in detail,\footnote{An example of of this kind in the translational sector was 
noted by Townsend \cite{townsend2}.} as we will focus on the worldvolume vacuum structure for 
which the two-derivative sector of the theory is sufficient.

In this specific example, one can show that on quantization there are no
supersymmetric vacua, and thus no quantum BPS kinks, 
since $(Q_2)^2$ is bounded from below by the 
scalar curvature ${\cal R}$ of $\widetilde{\cal M}$ which is clearly
positive. More precisely \cite{lsv}, $Q_2$ can be realized as the 
Dirac operator on $\widetilde{\cal M}$,
\be
 Q_2 = \frac{1}{\sqrt{2}} \si^j (-i \nabla_j)\ ,
\ee
and thus one finds
\be
 (Q_2)^2 = {\cal H}_{\rm SQM} = -\nabla^2 + \frac{1}{8}{\cal R}\ ,
       \label{Liech}
\ee
where $\nabla^2$ is the Laplacian on $\widetilde{\cal M}$.

Although we focused on one particular example, the mechanism
for SUSY enhancement exhibited above clearly generalizes readily to, for example, 
sigma models with target spaces which are (nontrivial) U(1) bundles $R$
over K\"ahler manifolds $K$ (e.g. regular Sasakian manifolds), where the soliton
profile in the transverse coordinate $z$ lies entirely within the $S^1$ fibre,
\bea
 S^1(z) \longrightarrow &R& \nonumber\\
                 &\downarrow& \nonumber\\
              &K&
\eea
and it would clearly be interesting to explore other generalizations. 
It is worth emphasizing here that a global feature
of this kind is ultimately what is responsible for enhancing the
supersymmetry on $\widetilde{\cal M}$. In other words, the existence of a globally
defined K\"ahler form on $\widetilde{\cal M}$, while not strictly proven through our local 
considerations, is ensured by an underlying geometric structure.
Note also that the nontriviality of the fibration is a necessary condition ensuring
that the $\{\xi_0,\ph_0\}$  zero modes are normalizable, i.e. that they are localized to
the kink.

Here we will note only one natural extension of the example 
above, which is directly relevant to our subsequent 
discussion of BPS walls in SQCD.
We can embed the \none $S^3$ model in a K\"ahler ${\cal N}=(2,2)$ 
sigma model with target space $T^*(S^3)$. The bosonic soliton solutions
persist, and depend only on the base $S^3$ coordinates. Consequently,
the bosonic moduli space is unchanged. However, the cotangent directions
supply an additional set of fermionic zero modes, so that the
bosonic and fermionic moduli are now paired one-to-two, and the 
reduced moduli space preserves the action of
four supercharges constructed as above with $\et$ reinterpreted as a
two component Majorana spinor. This in fact is crucial as
the system can then be lifted to a nonchiral theory in 1+1D with
${\cal N}=(2,2)$ supersymmetry and, more importantly, the natural
\ntwo theory in 2+1D on the worldvolume of a domain wall. 

In the next section, we will review the origin of this geometric structure
within the context of BPS domain walls in \none SQCD, and describe 
in some detail the worldvolume dynamics on the reduced moduli space.

\section{Domain Wall Moduli in \boldmath{\none} SQCD}

In the first part of this section we briefly review the arguments 
which determine the topology of the reduced $k$-wall moduli space in SU($N$) 
SQCD with $N_f=N$ flavors \cite{rsv}. We then present a more explicit 
construction for SU(2), exhibiting the enhancement 
of supersymmetry on the reduced moduli space, and describing 
how the imposition of a hierarchical structure for the quark mass matrix 
leads to a potential which lifts the flavor moduli.
This potential, at least for linear deformations, is geometrically 
the norm squared of a U(1) Killing vector. 

\subsection{The \boldmath{$k$}-wall moduli space and the CFIV index}

\none SQCD with $N_f=N$ flavors is obtained by adding $N$ chiral superfields,
$Q_{f}$ and $\widetilde {Q}^{\bar g}$ ($f,{\bar g}=1,\ldots,N$), transforming respectively in the
fundamental and anti-fundamental representations of the gauge group, to the fields
of \none SYM with gauge group SU($N$). This matter content will ensure that the
gauge symmetry is completely broken in any vacuum in which the matter fields
have a nonzero vacuum expectation value. Provided the mass gap is sufficiently large,
the gauge fields may then be integrated out, obtaining a low energy effective
description in terms of the meson moduli $M_f^{\bar g} = Q_f \widetilde {Q}^{\bar g}$.

The superpotential describing the resulting low energy dynamics is given by
\be
 {\cal W} = {\rm Tr}(\hat{m}M) + \la \left({\rm det}M-\La_N^{2N}\right), 
  \label{sp}
\ee
in terms of the meson matrix $M$, the dynamical scale $\La_N$, and
a Lagrange multiplier $\la$. The Lagrange multiplier is to be understood as a 
heavy classical field, for consistency with the nonrenormalization theorem, which
enforces a reduced form of the quantum constraint \cite{Seiberg},
\be
 {\rm det}M-B\widetilde{B} = \La_N^{2N}\ ,
\ee
containing in addition the baryon fields $B$ and $\widetilde{B}$.
These fields have been set to zero (their vacuum values) 
in (\ref{sp}) as they do not play a role in the wall configurations 
we will consider here.

An important constraint on the accessible parameter space is the requirement
that the vacua of the theory, and generic domain wall trajectories, lie
at weak coupling where the gauge modes, which have been integrated out, are indeed
heavy. This condition is satisfied if the quark mass matrix $\hat m$ is chosen
in a specific hierarchical form, and the choice which retains the maximal
global symmetry is given by
\be
 \hat{m} = {\rm diag}\{m, m ,\ldots, m, m_N \}\ , \;\;\;\;\;\;\; 
   \La_N \gg m_N \gg m\ . \label{hmm}
\ee
The vacua are then given by diagonal meson vacuum expectation values
(VEVs) with components (no summation over $i$),
\be
 \langle M_{i}^{i} \rangle_k = \left(\frac{m_N}{m}\right)^{1/N}\!\!\La_N^2 \om_N^k\,, 
 \quad \om_N^k={\rm e}^{2\pi k/N}\,, \quad
i=1,\ldots,N-1\,, \quad k=0,\ldots,N-1\ .\label{vac}
\ee
The vacua are weakly coupled if the hierarchy is 
sufficiently large: i.e. we require $m_N/m \gg e^N$.
If we restrict our attention to energy scales below $m_N$, the 
effective dynamical scale is $\La_{N-1}^{2N+1}=m_N\La_N^{2N}$.

For the specific problem of deducing the multiplicity of BPS walls
the need for a hierarchical mass matrix can be circumvented \cite{rsv}. 
This counting problem amounts to computing the CFIV index \cite{cfiv},
which is formally defined as the following trace, suitably regularized, 
over the Hilbert space with boundary conditions appropriate to a 
$k$-wall \cite{cfiv,cv},
\be
 \nu_k \equiv {\rm Tr}\, F (-1)^F\ ,
\ee
where $F$ is the fermion number operator. Note that only shortened
multiplets contribute. It will be useful to briefly recall two approaches 
to the computation of this index in the present context
(see \cite{rsv} for further details):

\bigskip
\noindent$\bullet\;${\it Hierarchical regime: counting permutations}
\bigskip

It is convenient to define dimensionless fields
$X=\hat{m}M(\mu\La_N^2)^{-1}$, with $\mu\equiv ({\rm det}\,\hat{m})^{1/N}$,
in terms of which the superpotential 
exhibits the maximal SU($N$) flavor symmetry,
\be
 {\cal W} = \mu\La_N^2 \left[ {\rm Tr}\,X + \la({\rm det}X - 1)\right],
   \label{spX}
\ee
while the hierarchical structure of the mass matrix is now visible
only in the rescaled K\"ahler potential. The superpotential depends only
on the eigenvalues  $\{\et_i\}$ of $X$,
\be 
 {\cal W} = \mu\La_N^2 \left[ \sum_{i=1}^N \et_i +
 \la\left(\prod_{i=1}^N \et_i - 1\right)\right],
\ee
which exhibits the vacua at the roots of unity, $\langle \et_i \rangle_k = \om_N^k$.
Specifying boundary conditions relevant for a $k$-wall, the trajectory
of each eigenvalue is characterized by its winding number $w(\et)$ which can take one
of two possible values: $w_1=k/N$ and $w_2=k/N-1$ (see also \cite{hiv}).
The Bogomol'nyi equations then ensure that $N-k$ of the eigenvalues carry winding
number $w_1$ and $k$ carry winding number $w_2$. It follows immediately 
that the wall multiplicity is given by the number of permutations of the eigenvalues
subject to these conditions, i.e.
\be
 \nu_k ~=~ \left( \begin{array}{c} N \\ k \end{array}\right)  ~=~ \frac{N!}{k!(N-k)!}\ . \label{nuk}
\ee
One observes that, since this construction depends only on constraints on the
$N$ eigenvalues, it can be applied consistently in the decoupling limit of 
the $N^{th}$ flavor.

\bigskip
\noindent$\bullet\;${\it Symmetric regime: quantizing moduli}
\bigskip

An alternative approach, developed in \cite{rsv}, involves noting that
the CFIV index can also be deduced from the Witten index Tr$(-1)^F$ \cite{Witten1} 
of the worldvolume theory on $\widetilde{\cal M}_k$. Thus it depends only
on the topology of the reduced moduli space of BPS walls.
These moduli are determined by the flavor symmetries broken by the wall and
parametrize a K\"ahler manifold.\footnote{The K\"ahler structure of the
reduced moduli space, which is not a priori imposed by SUSY, will be discussed
in more detail below.} In particular, it is only the
induced metric on this space which is sensitive to the precise
specification of quark masses; the topology is invariant. One 
then recalls that the CFIV index is independent of smooth diffeomorphisms of 
the K\"ahler potential \cite{cfiv}, and so we can restore its symmetry by such
a diffeomorphism if so desired. 

The result (\ref{nuk}) can then be understood via quantization of
the classical moduli space Lagrangian. In particular, it follows from the 
constraints on the eigenvalues that
the maximal flavor symmetry that the $k$-wall can preserve is
\be
 {\rm SU}(k) \times {\rm SU}(N-k) \times {\rm U}(1)\ ,
\ee
which is a subgroup of the full flavor symmetry SU($N$).
Consequently, taking care with discrete factors, there must be
localized Goldstone modes on the wall parametrizing the Grassmannian
coset \cite{rsv},
\be
 \widetilde{\cal M}_k = G(k,N) \equiv 
 \frac{{\rm U}(N)}{{\rm U}(k) \times {\rm U}(N-k)}\ . 
\ee
The CFIV index then reduces to the worldvolume Witten index of the 
supersymmetric Grassmannian sigma model, given by the Euler characteristic, with the result
\be
 \nu_k ~=~ \ch(G(k,N)) ~=~ \frac{N!}{k!(N-k)!}
\ee
for the multiplicity of $k$-walls, in the presence of a suitable infrared
regulator, consistent with the result above. 

This latter computation relies heavily on the invariance of the index under 
$D$-term deformations, in order to deform the theory to a symmetric
mass regime. We now wish to study this worldvolume theory in more detail
and consequently will need to consider more carefully the transition back to the
weakly coupled hierarchical mass regime.

\subsection{The SU(2) case and enhanced supersymmetry}

In order to provide a more explicit discussion of the resulting worldvolume dynamics
on the moduli space of BPS walls, we will limit our attention
in what follows to the simplest example with gauge group SU(2) and $N_f=2$ 
flavors. 

In addressing the full worldvolume dynamics, we are no longer
at liberty to perform diffeomorphisms of the K\"ahler metric, and so it will
be useful to introduce another dimensionless meson field $Z=M \La_N^{-2}$ in terms
of which the symmetry breaking induced by the hierarchical mass matrix is visible
within the superpotential. A convenient basis is then provided by the following
decomposition,
\be
 Z = U_{\al_2-\al_1} (Z_0 \Id + i Z_i \si^i) U_{\al_2-\al_1}, \;\;\;\;\;\;\;\;\;
   U_\al = \exp\left( \frac{i}{4} \al \si^3\right),
\ee
where the (axial) rotation angle is the relative phase of the two quark masses;
$m_k=|m_k|e^{i\al_k}$ for $k=1,2$. In this basis, the moduli space
constraint takes the form,
\be
 \sum_{a=0}^3 Z_a^2 = 1\ ,
\ee
and it describes a smooth complex submanifold of $\C^4$, known as the deformed 
conifold \cite{Candelas}. This manifold is symplectically equivalent to 
$T^*(S^3)$.

In studying the BPS wall spectrum, it will be convenient to first consider
the decoupling regime with hierarchical quark masses. 

\bigskip
\noindent$\bullet\;${\it The decoupling regime}
\bigskip

We first consider the regime where
\be
 \left|\frac{m_2}{m_1}\right| \gg 1\ ,
\ee
so that the second flavor can be integrated out. The superpotential 
can be written as follows
\be
 {\cal W} = e^{i\ga}|m_1|\La_2^2\left[ Z_+ +\left|\frac{m_2}{m_1}\right|  Z_- \right] 
  + \la \left( Z_+Z_- + Z_1^2 +Z_2^2 -1\right),
\ee
where $Z_{\pm} = Z_0 \pm i Z_3$ and $\ga=(\al_1+\al_2)/2$ is an overall phase. 
In the decoupling limit $Z_1$ and $Z_2$, since they
are sensitive to the heavy quark VEV, are set to zero 
$\langle Z_1 \rangle =\langle Z_2 \rangle =0$, and thus
the moduli space contracts to
\be
 Z_+Z_- = 1\ , \label{const1}
\ee
a submanifold which is locally $\R \times S^1$. Solving this
constraint directly, one recovers the Affleck-Dine-Seiberg superpotential
for the 1-flavor theory \cite{ADS}. We will instead proceed by restricting the fields
$Z_+$ and $Z_-$ to lie on the $S^1$ real section of (\ref{const1}), since this
contains the two vacua at 
$\langle Z_+ \rangle =\langle Z_- \rangle^{-1} = \pm \sqrt{|m_2/m_1|}$. Introducing an
angular coordinate $\th\in [0,\pi]$, we define
\be
 Z_+ = Z_-^{-1} = \sqrt{\left|\frac{m_2}{m_1}\right|} e^{\pm i\th}\ , \label{ambig}
\ee
which ensures that the physical meson field $M_{11}$ scales as $(\La_1^5/m_1)^{1/2}$
and thus remains finite in the decoupling limit. The
classical K\"ahler potential for $M_{11}$, which is reliable in
this hierarchical regime, also scales as $(\La_1^5/m_1)^{1/2}$.

The superpotential reduces to 
\be
 {\cal W} = 2e^{i\ga}\sqrt{|m_1|\La_1^5} \cos\th\ ,
\ee
which we recognize as equivalent, up to normalization, to the 
(real) superpotential of the $S^3$ model analyzed in Sec.~II. The Bogomol'nyi equation
takes the sine-Gordon form,
\be 
 \ptl_z \th = - 2|m_1| \sin\th\ ,
\ee
and thus the solution,
\be
 \th_{\rm sol}(z) = 2 {\rm arctan}\left(e^{-2|m_1|(z-z_0)}\right), \label{hiersol}
\ee
exhibits a single bosonic modulus $z_0$ corresponding to the soliton position. 
We also observe from the $\Z_2$ ambiguity in (\ref{ambig}) that there are two solutions \cite{Stwo}, 
consistent with the value of the index $\nu_1^{N=2}=2$.

Since there is no reduced moduli space for domain walls in this regime, we 
will not discuss the realization of supersymmetry explicitly. We note only
that the translational sector is described by a single free \none scalar multiplet 
in 2+1D. The results above will nonetheless provide a useful comparison to those
we will derive in the symmetric mass regime below.

\bigskip
\noindent$\bullet\;${\it The symmetric regime}
\bigskip

We would now like to consider this system outside the decoupling regime.
Although we will ultimately return to the controllable hierarchical mass
regime (albeit with $m_2$ finite), we will first abstract slightly and
consider what happens when we set the quark masses equal $m_1=m_2=\mu$. 
Although this puts the wall trajectory at strong coupling, it turns out that
the enhanced symmetry will still provide important constraints, and essentially
the only assumption we need to make is that the effective description in terms
of meson moduli is still valid. In practice, we do this simply to study the
kinematic structure of the resulting worldvolume superalgebra, and
we will partially remove the need for this assumption in the next subsection 
where we consider how the resulting picture
is modified on detuning the two quark masses.

To proceed, it is now convenient to write the superpotential in the
following form
\be
 {\cal W} = e^{i\ga}\La_2^2\left[ \bar{m} Z_0 + i \De m Z_3 \right] 
  + \la \left( \sum_{a=0}^3 Z_a^2-1\right),
\ee 
where the (real) mass parameters are
\be
 \bar m = |m_1| + |m_2|, \;\;\;\;\;\;\;\;\; \De m = |m_2| - |m_1|\ .
\ee

Setting $\De m=0$, we observe that the two vacua, $Z_0=\pm 1$, now lie at the poles of
the $S^3$ which forms the real section of the surface $\sum_{a=0}^3 Z_a^2=1$. 
Supersymmetry demands that the metric on this latter space be K\"ahler.
However its precise form is subject to quantum corrections and is not known
except in the asymptotic regime where $M_{f}^{\bar g} \gg \La_2^2$. Fortunately, one can show that not 
only the vacua but also the wall solutions  lie entirely within the $S^3$ section 
\cite{rsv} and we can ignore the metric structure of the cotangent directions. 
Moreover, when both mass terms are set to zero, the theory preserves an enhanced SU(2)$\times$SU(2) 
symmetry which demands that the induced metric on the base $S^3$ be the round one. We can introduce 
a suitable set of coordinates $\{\ph'^a\}$ for the surface $\sum_a Z_a^2 = 1$,
or a submanifold thereof,  which
makes the symmetry of this embedding manifest, and we denote the induced line element 
$d\Om_3'(\ph'^a)$.

Let us also introduce a second coordinate system for the 
$S^3$, $\{\ph^a\}$, given by the embedding into flat space,
with induced line element $d\Om_3(\ph^a)$. Note that one obtains the same result for the
embedding within the classical K\"ahler geometry ${\rm Tr}\sqrt{(\bar{Z}Z)}$. 
The relation between the two induced metrics  $d\Om_3'(\ph'^a)$ and $d\Om_3(\ph^a)$ is
nontrivial, and determined by the renormalization of the K\"ahler potential. However,
symmetry demands that we have
\be
 d\Om_3'(\ph'^a) = f(\ph^a) d\Om_3(\ph^a)\ ,
\ee
with a conformal factor $f(\ph^a)$, consistent with the isometries, which must be
nonsingular to preserve the known vacuum structure. Note that this is
a stronger constraint than would apply to the entire K\"ahler metric. We now see 
that, although $f(\ph^a)$ is unknown in general, it will enter the Bogomol'nyi equation for
BPS walls in such a form that it can be `removed' by a field-dependent rescaling of the
transverse spacetime coordinate to the wall. Such a rescaling will affect the wall profile,
but will not affect the symmetries of the system and will allow us to proceed with an
analysis of the kinematics. Thus, for this subsection, we will perform this
rescaling and set $\ph'^a=\ph^a$. The induced metric on the $S^3$, in spherical polar
coordinates  $\{\th,\xi,\ph\}$, 
then takes the form
\be
 ds^2_{\rm base} = \La_2^2 
 \left( d\th^2 + \sin^2\th \left( d\xi^2 + \sin^2\xi d\ph^2\right)
   \right), \label{2met}
\ee
where the normalization is fixed by the only dimensionful scale available,
the dynamical scale $\La_2$ of $N_f=2$ SQCD. When we turn the equal mass
perturbation back on, the corrections will be of order $\mu/\La_2$ which are subleading
in the light quark mass regime we consider here. In the next subsection, we will
consider unequal mass perturbations which will move the wall trajectory back
toward the weakly coupled region. 

We can now utilize the same coordinate system, $\{\th,\xi,\ph\}$, to rewrite the
superpotential, restricted to the $S^3$ real section, in the form,
\be
 {\cal W} = e^{i\ga}\mu \La_2^2 {\rm Tr}Z \longrightarrow 2e^{i\ga}\mu \La_2^2 \cos\th\ ,
\ee
which is once again equivalent, up to normalization, to the 
superpotential of the $S^3$ model analyzed in Sec.~II, and the 
superpotential in the hierarchical regime deduced above. In the latter context,
the use of the same notation for the angle $\th$ entering the superpotential 
is not accidental and will be justified later in this section.
The vacua lie at the poles $\th=0,\pi$, and the Bogomol'nyi equations
reduce to 
\be
 \ptl_z \th = - 2\mu \sin\th\,, \qquad \ptl_z \xi = \ptl_z \ph =0\ ,
\ee
which are naturally equivalent to Eq.~(\ref{bogeqn}), and are solved 
once again by the sine-Gordon soliton (\ref{sgsol}),
\be
 \th_{\rm sol}(z) = 2\arctan\left(e^{-2\mu(z-z_0)}\right), \;\;\;\;\;\;
   \xi_{\rm sol}=\xi_0,\;\;\;\;\;\;  \ph_{\rm sol}=\ph_0\ . 
\ee
We conclude that the bosonic moduli space is the same as that obtained within
the $S^3$ model, namely ${\cal M}^{\rm N=2} = {\R} \times {\C}{\rm P}^1$, which is
consistent with the general discussion above. 
Integrating over the wall profile, and reconstructing the spatial dependence
using Lorentz invariance, leads to the corresponding bosonic moduli 
space Lagrangian,
\be
  {\cal L}_{\rm bose} = \int d^3 x 
 \left[-T_1 + \frac{1}{2}\,T_1\, \partial_{\mu} z_0\, \partial^{\mu}z_0
 + \frac{1}{2}\,h_{ij} \ptl_\mu x^i \ptl^{\mu}x^j\right],
\ee
where $T_1=4\mu\La_2^2$ is the 1-wall tension, and $h_{ij}$ is the
metric on the moduli space, given by
\be
 ds^2_{{\cal M}} = T_1\, d z_0^2 +h_{ij}dx^idx^j = T_1\, d z_0^2 + 
 R_{\widetilde{\cal M}} \left( d\xi_0^2
 + \sin^2\xi_0 d\ph_0^2\right),
\ee
with 
\be
 R_{\widetilde{\cal M}} = \frac{\La_2^2}{\mu} 
    \label{Mscale}
\ee
the scale of the reduced moduli space.

We are now in a position to explore the realization of supersymmetry 
on the reduced moduli space. The first point to note, following
the comments at the end of Sec.~2, is that the present system has
twice as many fermions as the $S^3$ model considered earlier. The second
set of fermions arise from the cotangent directions of $T^*(S^3)$. We
can choose a basis where the complex fermions lying in the chiral
multiplet $Z$ decompose into two (real) sets, one $\psi_{1\al}$ 
the \none partner of the $S^3$ coordinates of the base, and 
the other $\psi_{2\al}$ the \none partner of the cotangent directions.
One then finds that a second set of fermionic zero modes arise
from $\psi_{21}$. The fermionic mode decomposition takes the
form
\bea
 \ps^a_{1\al} &=&\left[ \et_1^z
  \frac{\ptl \ph^a_{\rm sol}}{\ptl z_{0}}+  \et_1^i 
  \frac{\ptl \ph^a_{\rm sol}}{\ptl x^i} \right]
   \left(\begin{array}{c} 0 \\ 1 \end{array}\right)_{\!\al} + \mbox{nonzero modes}\,,
    \nonumber \\[1mm]
 \ps^a_{2\al} &=& \left[ \et_2^z
  \frac{\ptl \ph^a_{\rm sol}}{\ptl z_{0}}+\et_2^i 
  \frac{\ptl \ph^a_{\rm sol}}{\ptl x^i} \right]
   \left(\begin{array}{c} 1 \\ 0 \end{array}\right)_{\!\al} + \mbox{nonzero modes}\ ,
\eea
where $\{\et_A^i\}$ are two sets of fermionic operators satisfying
\be
  \{\et^z_A,\et^z_B\} = \frac{1}{T_{1}}\,\delta_{AB}\,,\qquad\{\et^i_A,\et^j_B\} = h^{ij}\delta_{AB}\ ,
\ee
where $h_{ij}$ is the reduced moduli space metric.
Thus we now find in full a one-to-two matching between the
number of bosonic versus fermionic zero modes. It is important that
since the worldvolume is now 2+1-dimensional, this matching condition
is a requirement of Lorentz invariance -- a constraint that was
not present in our earlier discussion of 1+1D kinks.

With this constraint in mind, it is convenient to combine 
the fermionic moduli $\et_A$ into a two-component
spinor $\et=(\et_1,\et_2)$. The center-of-mass sector now comprises a real scalar
$z_0$ and a Majorana spinor $\et^z$, which is sufficient to
compose a scalar multiplet of \none SUSY in 2+1D. This is the sector
of the theory generated by spontaneous breaking of translational
invariance and the two broken supercharges, since the state is
1/2-BPS. 

The reduced moduli space ${\C}{\rm P}^1$ is K\"ahler, and so
from the discussion of Sec.~2, we would anticipate some `enhancement'
of supersymmetry in this sector. Indeed, it is clear that essentially
the same construction as before, now augmented with two-component
spinors $\et^i$, will lead to the dynamics admitting \ntwo 
supersymmetry in 2+1D, or four supercharges, only two of which
can be identified with the unbroken generators of the bulk
superalgebra. This conclusion has important consequences, as 
this theory can be shown to have two supersymmetric vacua (at least
in 1+1D or less), in contrast to the `chiral' theory which
was realized in the $S^3$ model.

In preparation for the following subsection, it will be useful to describe  
explicitly the construction of the supercharges. To this end, we will
compactify the theory on a 2-torus of radius $R$, and consider the 
${\cal N}$=(2,2) superalgebra in 1+1D:
\bea
 \{Q_\al, Q^\dagger_\beta\} &=& 2(\ga^\mu\ga^0)_{\al\beta}P_\mu \, , \nonumber\\[1mm]
  \{Q_\al, Q_\beta\} &=& 2i(\ga^5\ga^0)_{\al\beta}\bar{\cal Z}\, ,  \nonumber\\[1mm]
  \{Q^\dagger_\al, Q^\dagger_\beta\} &=& 2i(\ga^5\ga^0)_{\al\beta} {\cal Z}\,,
\eea
choosing the $\ga$--matrices as follows
\be
 \ga^0=\si_2\,, \quad \ga^1 = i\si_3\,, \quad \ga^5=\ga^0\ga^1 = -\si_1\ .
\ee
We can now rotate to a Majorana basis 
$Q_{\al}=e^{-i\ga/2}(Q_\al^1 + i Q_\al^2)/\sqrt{2}$ within which
\be
 \{Q_\al^i, Q_\beta^j\} = 2\de^{ij}(\ga^\mu\ga^0)_{\al\beta}P_\mu + 2 i 
    (\ga^5\ga^0)_{\al\beta}
                                |{\cal Z}| \si_3^{ij}\ ,  \label{realform}
\ee
where $\ga\equiv{\rm arg}({\cal Z})$, and takes the form $\ga=(\al_1+\al_2)/2$ in the present case.
In the rest frame,
\bea
 &&(Q_1^1)^2 = (Q_2^2)^2 = M + |{\cal Z}|\,, \nonumber\\[1mm]
&& (Q_1^2)^2 = (Q_2^1)^2 = M - |{\cal Z}|\, ,
\eea
where $M=T_1 R^2$ in terms of the wall tension. Thus, we see that for a BPS wall
configuration $Q_1^2$ and $Q_2^1$ are the unbroken supercharges which will be
realized within the worldvolume theory.

To compute these supercharges in terms of the moduli we recall that
for a Wess-Zumino model, as we have here, the complex supercharge is given by
\be
 Q = \int dz \left( g_{ab} \ga^\mu \ptl_\mu \ph^a \ga^0 \ps^b + i \ptl_{\bar b} \bar{\cal W}
    \ga^0 \ps^{* b}\right).
\ee
To move to the Majorana basis, we decompose $\ps = e^{-i\ga/2}(\ps^1 + i \ps^2)/\sqrt{2}$, 
and obtain
\be
 Q^1 = \int dz \left( \begin{array}{cc}
                         g_{ab} \dot{\ph}^a & g_{ab}\ptl_z \ph^a\! +\! e^{i\ga}\ptl_{\bar b} 
                           \bar{\cal W}_0 \\[1mm]
                          g_{ab}\ptl_z \ph^a \!-\! e^{i\ga}\ptl_{\bar b} 
                     \bar {\cal W}_0 & g_{ab} \dot{\ph}^a
                      \end{array} \right) \ps^{1b},
\ee
with $Q^2_{\al}$ given by a similar expression in terms of $\ps^2$.
Inserting the solutions for the fermionic zero modes one obtains,
\be
 Q_\al \equiv (Q^1_2,Q_1^2) = T_1 \dot{z}_0 \et^z_\al+ h_{ij} \dot{x}^i \et^i_{\al}\ ,
\ee
where we have combined the two unbroken charges into a spinor, using the
corresponding fermionic zero modes $\et_\al$. This is recognizable as a spinor
analogue of the unbroken supercharge within the $S^3$ model. If we now drop
the translational zero modes, and restrict $Q_{\al}$ to the reduced moduli 
space, with $x^i=\{\xi_0,\ph_0\}$, then we discover that there is a second unbroken 
spinor supercharge, existing by virtue of the complex structure $J$ 
associated with $\widetilde{\cal M}=S^2$, introduced earlier in (\ref{compstruc}). 
We can then form a complex
spinor charge ${\cal Q}_\al^I$
\bea
 {\cal Q}_\al^I \equiv \left\{ \begin{array}{l} 
                               Q_\al = h_{ij} \dot{x}^i \et_\al^j\,, \\[1mm]
                  \widetilde{Q}_\al = h_{ij} J^j_k \dot{x}^i \et_\al^k\,,
                           \end{array} \right.
\eea
and these charges satisfy the algebra of \nfour SQM or more importantly, when lifted
back to 2+1D, the \ntwo superalgebra.

The worldvolume theory is then an \ntwo $\cp^1$ sigma model and, as noted above,
the Witten index for this theory is equal to two, consistent with our counting of domain
walls. Therefore, within this system, at least when compactified to 
1+1D or below, there are indeed two quantum vacua, and thus
two BPS walls. This worldvolume structure also has important
consequences for worldvolume BPS solitons, a subject that
we turn to in the next section.

The crucial distinction to be made here with the 
\none algebra arising for kinks in the $S^3$ model
is that with two-component fermions the model has an additional
potential term associated with the Riemann tensor, 
\be
 \De V = - \frac{1}{12} \, R_{ijkl} \,\bar{\et}^i \et^j \bar{\et}^k \et^l\ ,
\ee
with $\bar{\eta}=\et^T\ga_0$, which precisely cancels
the zero-point curvature term in (\ref{Liech}) in the quantum action of the
unbroken supercharges on the ground states. One way to understand this
is to recall\footnote{Similar issues arise in comparing the spectrum of dyons of magnetic
charge two in gauge theories with \ntwo \cite{gauntlett93} and \nfour supersymmetry \cite{sen}.} 
that, while the one-component worldvolume supercharges which arise in the \none $S^3$ model 
are realized quantum mechanically 
in terms of the Dirac operator, or alternatively (anti-) holomorphic 
(or Dolbeault) exterior derivatives \cite{gauntlett92}, on the reduced moduli space,
\be
  ({\cal Q}, {\cal Q}^*) \leftrightarrow (\bar{\ptl}^{\dagger},\bar{\ptl})\ ,
\ee 
the spinor supercharges arising in the \ntwo $T^*(S^3)$ model are realized in 
terms of (de Rham) exterior derivatives \cite{Witten1},
\be
 (Q,Q^{\dagger}) \leftrightarrow (d, d^{*})\ .
\ee 
The supersymmetric vacua in the latter case correspond to 
normalizable harmonic forms, of which there are two
for $S^2$ corresponding to the Betti numbers $b_0=b_2=1$. However, 
supersymmetric vacua of the \none theory would be normalizable
{\it holomorphic} harmonic forms on the same manifold. The presence of such forms on
a K\"ahler manifold, which would necessarily have anti-holomorphic partners, is 
forbidden by the uniqueness of harmonic forms in each de Rham cohomology class.

\bigskip
\noindent$\bullet\;${\it Quark mass splitting and a potential on the moduli space}
\bigskip

In the preceding discussion, we abstracted slightly in ignoring the
deformation imposed by considering a hierarchical mass matrix for the
quarks. As noted above, this choice is enforced if we wish to
retain a weakly coupled description of the vacua between which the
wall interpolates. In this subsection, we rectify this by turning
on this deformation and demonstrating that the effect on the reduced
moduli space is, at linear order, to introduce a new potential given by the 
norm squared of a U(1) Killing vector. Such a potential is 
naturally associated with the fact that turning
on the quark mass difference, $\De m \equiv |m_1| - |m_2|$,
breaks the nonabelian part of the global symmetry from SU(2)$\rightarrow$U(1).
An important feature of this particular deformation on the worldvolume is that it preserves 
the enhanced \ntwo SUSY \cite{agf}.

Using the same coordinate system as above, and restricting once again to the
real section, we can write the superpotential in the form
\be
 {\cal W} = {\cal W}_0 + i \De {\cal W} = e^{i\ga} \bar{m} \La_2^2[ 
         \cos\th + i \ep \sin\th \cos\xi]\ ,
\ee
where the (real) deformation parameter is
\be
 \ep \equiv \frac{\De m}{\bar{m}}\ .
\ee

Rather than study the exact wall solutions within this system, we will
consider the impact at leading order in $\ep$ on 
the moduli space dynamics valid at $\De m=0$. Working to linear order in $\ep$, 
it is consistent to make use of the unperturbed wall solution in 
constructing the worldvolume supercharges. The deformation is then apparent 
in the presence of a correction term,
\be
 \De Q^1 = \int dz \left( \begin{array}{cc}
                         0 & e^{i\ga}\ptl_b \De \bar{\cal W} \\[1mm]
                         - e^{i\ga}\ptl_b \De \bar{\cal W} & 0
                      \end{array} \right) \ps^{2b}\ ,
\ee
with the correction to $Q^2_{\al}$ given by a similar expression. 

It is now clear that at linear order in $\ep$ the broken
supercharges $Q_1^1$ and $Q_2^2$ are not corrected on setting the nonzero modes to
zero. This is consistent with the fact that corrections to the central charge
start at ${\cal O}(\ep^2)$,
\be
 {\cal Z} = 2e^{i\ga}\bar{m} \La_2^2\Big( 1 - \frac{1}{2} \,\ep^2  +\cdots \Big).
\ee

In contrast, the unbroken supercharges are corrected, and evaluating them using the
zeroth order Bogomol'nyi equations, we find
\bea
 Q^1_2 &=& T_1 \dot{z}_0 \et_1^z+ h_{ij} \dot{x}^i \et_1^j + \pi \ep \La_2^2\sin\xi_0 \et_2^\xi, 
          \nonumber\\[1mm]
 Q^2_1 &=& T_1 \dot{z}_0 \et_2^z+h_{ij} \dot{x}^i \et_2^j - \pi \ep \La_2^2\sin\xi_0 \et_1^\xi.
\eea
The relative sign for the perturbations to $Q^1_2$ and $Q^2_1$ 
ensures that $\{Q^1_2,Q^2_1\}=0$ in the rest frame as required.

We would now like to determine whether or not this linearized deformation 
has preserved the additional supersymmetry, associated with the complex structure
on the reduced moduli space. In fact we can verify this explicitly.
To proceed, let us drop the decoupled translational mode as above 
and relabel the supercharges acting on the reduced moduli space as follows
\bea
 Q^1_2 \longrightarrow Q^1_L &=& 
   2\,\frac{\La_2^2}{\bar{m}}\Big[\dot{\xi}_0\et_1^\xi + \sin^2\xi_0 \dot{\ph}_0\et_1^\ph 
    + \frac{1}{2}\,\pi\, \ep \bar{m} \sin\xi_0 \et_2^\xi \Big],\nonumber\\[1mm]
 Q^2_1 \longrightarrow Q^2_R &=& 
  2\,\frac{\La_2^2}{\bar{m}}\Big[\dot{\xi}_0\et_2^\xi + \sin^2\xi_0 \dot{\ph}_0\et_2^\ph 
     - \frac{1}{2}\,\pi \,\ep \bar{m}\sin\xi_0 \et_1^\xi \Big]. 
\eea
Remarkably enough one can write down a second set of supercharges leading to
the same Hamiltonian,
\bea
 Q^2_L &=& 2\,\frac{\La_2^2}{\bar{m}}\sin\xi_0 \Big[ \dot{\xi}_0\et_1^\ph - \dot{\ph}_0\et_1^\xi 
     + \frac{1}{2}\,\pi\, \ep\bar{m} \sin\xi_0 \et_2^\ph \Big],\nonumber\\[1mm]
 Q^1_R &=& 2\frac{\La_2^2}{\bar{m}}\sin\xi_0 \Big[ \dot{\xi}_0\et_2^\ph 
 - \dot{\ph}_0\et_2^\xi
      - \frac{1}{2}\,\pi\, \ep\bar{m} \sin\xi_0 \et_1^\ph \Big],
\eea
and one can verify that $\{Q_L^1,Q_L^2\}=\{Q_R^1,Q_R^2\}=0$, and 
$(Q^2_L)^2 = (Q_R^1)^2 = {\cal H}$. It follows that we can build complex
combinations of the form, $Q_L=(Q_L^1 + iQ_L^2)/2$ and $Q_R=(Q_R^1 + iQ_R^2)/2$, i.e.
\bea
 Q_L &=& \frac{\La_2^2}{\bar{m}}
 \Big[\big(\dot{\xi}_0\et_1^\xi + \sin^2\xi_0 \dot{\ph}_0\et_1^\ph\big)
    + i \sin\xi_0 \big(  \dot{\xi}_0\et_1^\ph - \dot{\ph}_0\et_1^\xi\big) \Big. \nonumber \\[1mm]
 && \qquad
   +\Big.\frac{1}{2}\,i\pi \ep\bar{m} \sin^2\xi_0 \et_2^\ph 
  + \frac{1}{2}\,\pi \ep\bar{m} \sin\xi_0 \et_2^\xi\Big],
        \label{Qcomp}
\eea
such that
\be
 \{Q_L, \bar{Q}_L\} = \{Q_R, \bar{Q}_R\} = {\cal H}\ ,
\ee
with the other anticommutators vanishing in the absence of central charges.

This structure is of course not accidental. We can make the underlying complex
structure manifest, by introducing complex coordinates associated with the
stereographic projection. If, as in (\ref{ccoord}), we define:
\be
 w = e^{i\ph_0}\tan\frac{\xi_0}{2}\ , \label{compcoord}
\ee
the corresponding map for the fermions is given by
\be
 \ps_{L} = \frac{1}{2} \,\sec^2\frac{\xi_0}{2} e^{i\ph_0} \Big( \et_1^{\xi} 
 + i \sin\xi_0 \et_1^{\ph}\Big),
\ee
with a similar relation for $\ps_R$ in terms of $\et_2^\xi$ and $\et_2^{\ph}$.
With these redefinitions, the somewhat lengthy expressions above for $Q_L$ and $Q_R$
take the simple form
\bea
 Q_L &=& h_{w\bar{w}} \left[ \dot{\bar w} \ps_L + \pi\ep\bar{m} \bar{w} \ps_R \right], 
 \nonumber\\[1mm]
 Q_R &=& h_{w\bar{w}} \left[ \dot{\bar w} \ps_R - \pi\ep \bar{m}\bar{w} \ps_L \right], 
\eea
with the remaining supercharges given by $\bar{Q}_L$ and $\bar{Q}_R$. The 
Fubini-Study metric is
\be
 h_{\bar{w} w}  =  4\,\frac{\La_2^2}{\bar{m}}\frac{1}{(1+ |w|^2)^2} 
= \frac{2R_{\widetilde{\cal M}}}{(1+ |w|^2)^2}\ .
\ee
This is precisely the structure expected for a deformation by a Killing vector proportional
to a `twisted' \cite{ghr,hh,dorey98} or `real' mass term in 1+1D or 2+1D respectively, thus 
preserving \ntwo SUSY. In fact, since $\ep$ is a real parameter, we see that 
this deformation is most directly interpreted as a `real'
mass term in 2+1D, as one would expect for the worldvolume theory of a 
wall in 3+1D. In this context $\ps_L$ and $\ps_R$ are then the upper 
and lower components respectively of a complex spinor.

We have focused on the impact of this deformation on the supercharges, since
we were working to linear order and making use of the undeformed soliton
solution. This deformation is visible at the bosonic level as a potential
given by the norm squared of a U(1) Killing vector $G = G^i \ptl_i$ for
rotations in $\ph_0$,
\be
 G^i = \frac{1}{2}\pi \De m \de^{i\ph_0}\ .
\ee
However, this contribution is of second order in the 
perturbation. Formally, we obtain
\be
 (Q^1_2)^2 = (Q_2^1)^2 = {\cal H} = \frac{1}{2}\, h_{ij} \dot{x}^i\dot{x}^j 
 + \frac{1}{4}\, \bar{m} \La_2^2 (\pi \ep)^2\sin^2\xi_0\,,
\ee
and thus the induced potential is of the form
\be
 \De V = \frac{1}{2} h_{ij} G^i G^j = \frac{1}{4} \,\bar{m}\La_2^2
\, (\pi \ep)^2\sin^2\xi_0\ . \label{potl}
\ee

Strictly speaking we have not verified that this structure indeed persists 
at second order in $\ep$. The difficulty is that in perturbing away from the 
symmetric point, we lose any semblance of control over the induced metric
on the moduli space, and one cannot rule out singularities arising in the
truncation to the real section -- these would most likely take the form
of cusps appearing at the vacua. This hinders a purely bosonic construction
via completing the square in the Hamiltonian \'a la Bogomol'nyi. Nevertheless,
we would like to emphasize here that the picture one obtains from
(\ref{potl}) is entirely consistent with the results we obtained earlier in
the opposite (hierarchical) limit in which $|m_2/m_1|\rightarrow \infty$.
In particular, the potential implies that the vacua lie at $\xi_0=0,\pi$.
From the polar coordinatization of the real section, we see that this
contracts the moduli space as follows:
\be
 Z_{\pm} = Z_0 \pm i Z_3 = e^{\pm i\th}\,, \;\;\;\;\; Z_1=Z_2=0\ .
\ee
This is entirely consistent with the behavior of the wall solutions
we observed in the hierarchical limit, accounting for the fact that here 
$|m_1/m_2|=1+{\cal O}(\ep)$. This consistency suggests that although
we have only considered the perturbation at linear order, the resulting physical
picture is valid more generally.

In concluding this section,  we will comment briefly on some subtleties that arise in
extending these arguments to higher $N$. Firstly, since the reduced moduli space for 1-walls,
$\cp^{N-1}$, can always be embedded within a suitably oriented real section of the meson moduli space
det$M=\La_N^{2N}$, it seems clear that the one-to-two pairing between bosonic and fermionic zero modes
will hold more generally. This ensures that multiplets when realized in terms of \none SUSY are
necessarily reducible, although this structure may of course be lifted once one goes beyond 
the two derivative level. With this matching, \ntwo SUSY would follow immediately given 
a K\"ahler metric on the reduced moduli space. It is this latter property, namely that the 
induced geometry is in fact {\it globally} K\"ahler, which appears difficult to prove in generality. 
In the symmetric mass regime, it of course follows directly from the
construction of the moduli space as a K\"ahler quotient. However, this regime is not weakly 
coupled and in the tractable hierarchical mass regime one loses the isometry constraints on the induced
metric. Nonetheless, the explicit construction in the SU(2) case is certainly suggestive 
that supersymmetry enhancement also arises for generic $N$ and $\widetilde{\cal M}_k$.

It is worth noting that this conclusion is rather novel when
the K\"ahler structure is not imposed by the residual supersymmetry of the BPS state. 
For point-like or string-like solitons one has
additional freedom through the possibility of realizing SUSY using
one-component fermions. Indeed, this is the
conventional manner in which worldline theories for e.g. lumps in 
K\"ahler sigma models, and monopoles in \ntwo SYM, get around the
apparent contradiction of being 1/2-BPS states while at the same
time having a K\"ahler, or respectively hyper-K\"ahler, moduli 
space \cite{gauntlett92,gauntlett93}.
The worldvolume theories in question can be thought of
as reductions of (0,2) and (0,4) sigma models in 1+1D, and 
this structure can be understood from the fact
that the same bosonic moduli space arises in theories with twice as
much supersymmetry, namely hyperK\"ahler sigma models and \nfour SYM,
where the additional fermionic zero modes restore the `nonchiral' 
structure to the worldvolume superalgebra.

\section{On 1/4-BPS Wall Intersections}

In this section, we will turn our attention to a second set of BPS
configurations present in \none SQCD. An inspection of the \none 
super-translation algebra in 3+1D shows that it admits central charges
supported by domain walls and also string-like sources (see Appendix). 
The corresponding charges
transform in the (0,1) and (1/2,1/2) Lorentz representations respectively.  
The SQCD theories considered here are not expected to exhibit BPS string 
solutions, but one has the possibility of forming (1/4-BPS) intersections 
or junctions of domain walls supported by both wall and 
string charges. One class of 1/4-BPS junctions arises from a multi-spoke
configuration of $N$ domain walls in theories with $N$ degenerate vacua. 
The 1/4-BPS criterion amounts to the statement that the superpotential evaluated
on a path through each wall surrounding the junction traces out a 
closed polygon \cite{gt99,dwj1,dwj2,dwj3}. However, for SQCD, the existence of a
degenerate spectrum of $k$-walls \cite{Stwo,rsv}, presents the 
possibility of forming a novel class of domain wall junction configurations 
consisting of only {\it two} walls.\footnote{A similar class
of string junctions consisting of just two strings was studied recently
\cite{vortex} in the context of gauge theories with eight supercharges.
From the standpoint of the bulk theory such string
junctions turn out to represent `confined monopoles' in the Higgs phase.} 
It is these configurations that we
will study in this section, first from the bulk perspective, and then from the 
worldvolume point of view of the constituent walls. We will generally
restrict our attention to gauge group SU(2) with $N_f=2$, and make use
of the worldvolume theory constructed in the previous section.

Before describing the explicit construction, we recall some well-known
(and some less well-known) features of the kinematics. To this end,
it is convenient to represent the superalgebra in 2+1D, which we can do by 
lifting the corresponding discussion of Sect.~3, phrased in
a 1+1D language appropriate to domain walls, to 2+1D compactified on a circle of 
radius $L$. Using the same notation, with the identification $\ga^2=i\ga^5$, 
we can extend (\ref{realform}) as follows \cite{A7,dwj2}:
\be 
  \{Q_\al^i, Q_\beta^j\} = 2(\ga^\mu\ga^0)_{\al\beta}\de^{ij}P_\mu 
   + 2i(\ga^0)_{\al\beta} \ep^{ij} {\cal Z}_{\rm S} 
   + 2(\ga^2\ga^0)_{\al\beta} (\si_3)^{ij} (|{\cal Z}_{\rm W}|L) \ ,
\ee
which includes, in addition to the lift of the kink (or wall) charge, 
denoted ${\cal Z}_{\rm W}L$, a new (real) charge, ${\cal Z}_{\rm S}$,
associated with localized objects in 2+1D -- which we have
taken to be positive to simplify the discussion.  In the rest frame, 
it is sufficient to focus on
the sector of the two unbroken supercharges in the background of a BPS domain
wall, namely $Q_2^1$ and $Q_1^2$, which we relabel as $Q'_i$ for $i=1,2$ respectively.
The rest-frame algebra in this subsector takes the form,
\be
 \{Q'_i,Q'_j\} = 2\de_{ij}(M-|{\cal Z}_{\rm W}|L) - 2(\si^1)_{ij}{\cal Z}_{\rm S}\ ,
\ee
from which we observe that, in a background with both central charges nonzero, only
one of these supercharges can annihilate the state, and the Bogomol'nyi bound takes
the form
\be
 M > |{\cal Z}_{\rm W}|L + {\cal Z}_{\rm S}\ . \label{bnd3}
\ee
1/4-BPS junction configurations are required to saturate this bound.

When we lift this picture one further dimension to 3+1D, an additional subtlety arises
from the fact that the charge ${\cal Z}_{\rm S}$, now associated with string-like
sources, transforms as a vector and is not algebraically independent of the
momentum, i.e. in 3+1D,
\be
 \{{Q}_\al, \bar{Q}_{\dot\al}\} = 
 2 (\ga^{\mu})_{\al\dot\al}(P_\mu + {\cal Z}^{\rm S}_\mu)\ . \label{sc}
\ee
Ignoring the wall charges for now, if we 
orient the string-like source in the $x_3$ direction, we see that the
BPS bound takes a somewhat unusual form
\be
 T^{(S)} \geq P_3 + {\cal Z}^{\rm S}_3 \geq {\cal Z}^{\rm S}_3\ , \label{stringcharge}
\ee
where the second relation follows on noting that for configurations which saturate 
the bound (\ref{bnd3}) in 2+1D, the allowed boost in the $x_3$ direction is `chiral',
namely in the current basis (with positive ${\cal Z}_{\rm S}$) 
$P_3$ is required to be strictly positive \cite{gt99}.
The crucial point here is that although the central charge is not algebraically independent
of the momentum, it is {\it dynamically} distinguished by the existence of an alternate
means of identifying $P_{\mu}$ via the conserved, and symmetric, energy-momentum
tensor. One can of course pick the `rest' frame $P_3=0$ to recover a more standard
form of the Bogomol'nyi bound as discussed in \cite{gt99,dwj1,dwj2}, but one can alternatively
`boost' the BPS soliton (see Fig.~2). The additional invariant which accounts for this
is $P_\mu{\cal Z}^\mu$, which in the present coordinate system reduces to $P_3$. 

This example illustrates the general point that the full spectrum
of central charges is not always obtained by lifting the algebra to the maximal allowed dimension,
and then matching the full number of components in the anticommutator of supercharges, minus the 
momenta, with the allowed set of tensor central charges. The reason is that 
not all vectorial charges can be absorbed into the momenta since, although they are not
algebraically independent, they are dynamically distinguished. We provide a 
discussion of the central charge structure along these lines in the Appendix.

Returning to the bound (\ref{stringcharge}) in the present context, the
fact that the only configurations currently known which
saturate this bound in \none theories are wall-junction configurations may partially be
explained by considering the number of localized zero modes, as discussed in Section~II. 
The one broken supercharge furnishes the junction with a single fermionic
zero mode. This is paired with a single bosonic zero mode whose
origin is best understood by viewing the junction as a kink-soliton
on the wall worldvolume. The bosonic zero mode then arises from 
the breaking of translational invariance along the wall. A second 
translational zero mode, associated with the position of 
the junction in the orthogonal
direction is not localized as it corresponds to a shift of 
wall itself. In this sense the worldvolume structure of the 
junction is quite distinct from a localized source such as a
vortex.

\begin{figure}
\includegraphics[width=6cm]{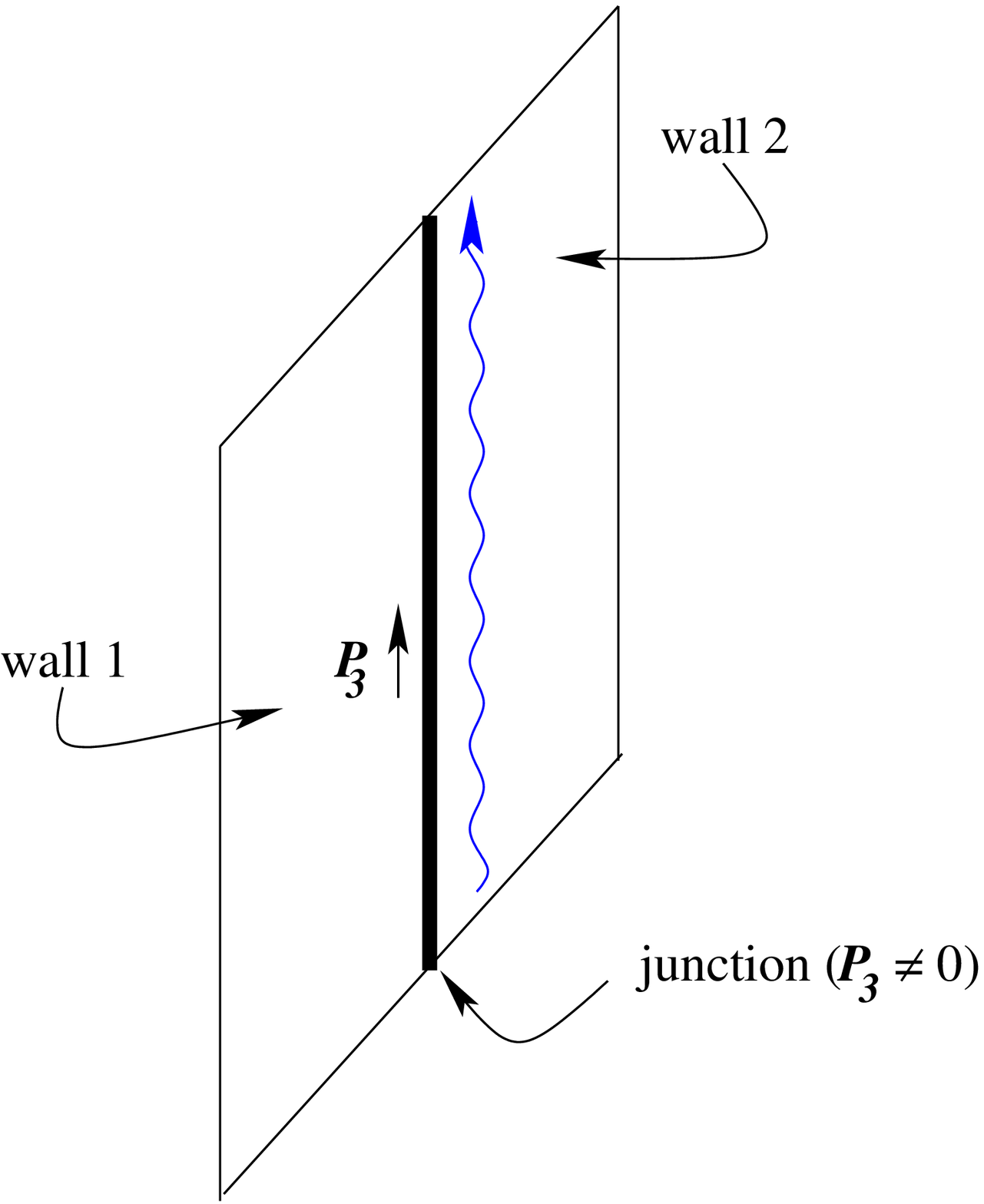}
\vspace*{0.3cm}
 \newcaption{\footnotesize A schematic representation of a `boosted'
BPS junction, i.e. a junction superposed with a wave of momentum $P_3$.}
\end{figure}

In 3+1D this structure has a natural interpretation in terms of the
extension of the zero modes to worldvolume fields in 1+1D. i.e. we can now
complete the single fermionic zero mode to a chiral fermion, which
we can choose to be left moving. In the bosonic sector 
the single translational zero
mode is completed to a bosonic field, which we can 
decompose into a left and a right-mover.
Only the left-mover will be paired with the fermionic zero
mode under the residual chiral (0,1) worldsheet supersymmetry \cite{gt99}.
Now, when we turn on $P_3$, we can interpret the
resulting junction which remains BPS as equivalent to
the `bare' $P_3=0$ junction superposed with a left moving 
wave of momentum $P_3$. This configuration is illustrated schematically in
Fig.~2.

With these preliminaries, we now return to the specific case of interest,
namely 2-wall junctions in \none SQCD, and consider
these solutions first in the hierarchical quark mass limit
with only one light flavor.

\subsection{Junction tension for $N_f=1$}

We first consider the hierarchical regime for gauge group SU(2), 
and integrate out the second flavor as in the corresponding discussion of
Sect.~3. Since we take the decoupling limit directly, and thus
solve the constraint $Z_+Z_-=1$ explicitly for $Z_+$, it is useful to
introduce another dimensionless field $Y$ in the form 
$Y=\sqrt{M_{11}}(\La_1^{5}m_1^{-1})^{-1/4} = \sqrt{Z_{+}}(m_2 m^{-1}_1)^{-1/4}$,
such that after decoupling
\be
 {\cal W} = \sqrt{m_1\La_1^5}\,\left(Y^2 + Y^{-2}\right), \;\;\;\;\;\;\;
   {\rm and}\;\;\;\;\; {\cal K} = \sqrt{\La_1^5m_1^{-1}}\,\bar{Y}Y\ .  
\ee
Provided we take $m_1\ll\Lambda_1$, the vacua $\langle Y^2 \rangle = \pm 1$
lie at weak coupling, and one can construct the two BPS wall configurations
we exhibited in (\ref{hiersol}) (first obtained in \cite{Stwo}), which
we reproduce here in the form (with $\tilde{\th}\in [-\pi,\pi]$),
\be
 Y^2_{\rm wall} = e^{i\tilde\th(x)}\,, \;\;\;\;\;\; 
  \tilde\th_{\rm sol}(x) = \pm 2 \,{\rm arctan}\left(e^{-2|m_1|(z-z_0)}\right)\,. 
   \label{wtraj}
\ee
The labeling of the two walls, $W_+$ and $W_-$, reflects whether
the phase of $Y^2$ interpolates between the two (real) vacua
via the upper or lower half-plane. The corresponding trajectories 
are illustrated in Fig.~3.

Having two degenerate walls, we can contemplate the possibility of a
2-wall junction in the form illustrated in Fig.~3. We choose
coordinates so that the walls interpolate from $k=0$ at $x=-\infty$ to $k=1$ 
at $x=+\infty$, and position the walls at $x_0=0$. The spatial worldvolume
dimension of the walls transverse to the junction 
will be denoted $y$, with the junction located at $y_0=0$.

Qualitatively, we see that at large $|y|$, remote 
from the junction, the field profiles are essentially those 
for the wall trajectories (\ref{wtraj}), i.e. 
$W_\pm$ for $y$ positive or negative. However, the evolution 
in $y$ must interpolate smoothly between $W_+$ and $W_-$. 
A (presumably rapid) transition necessarily occurs near $x=0,y=0$ 
where the junction is located. In particular,
such a smooth interpolation means that near $y=0$
our $x$ trajectory necessarily runs through the shaded
domain of small $Y$ shown on the right of Fig. 3, implying that strong dynamics must become 
important.

\begin{figure}[t]
\includegraphics[width=8cm]{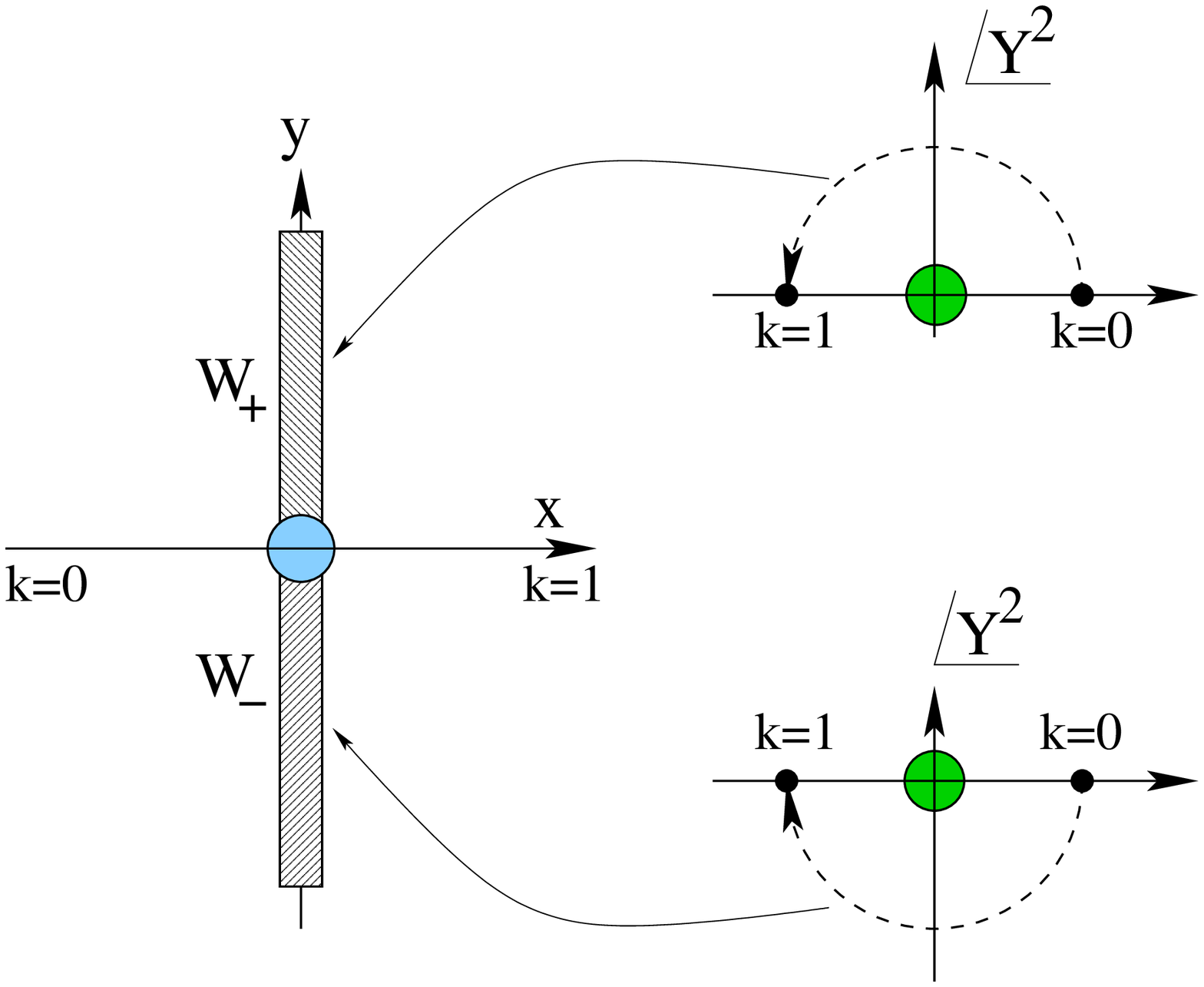}
\vspace*{0.3cm}
 \newcaption{\footnotesize The wall junction geometry, 
indicating the field profiles
in the $W_+$ and $W_-$ components. Note that, by continuity, the fields
 near the junction must pass through the strong-coupling regions near $Y=0$,
 where the low energy description breaks down.}
\label{ff4} 
\end{figure}

We can understand this more clearly by studying the Bogomol'nyi equation
which follows either by minimizing the energy, or equivalently
requiring a configuration preserving one of the four supercharges
in the \none algebra. Introducing the complex spatial coordinate $z=x+iy$, 
the equation can be written as \cite{CS,dwj3}
\be
 g_{\bar{Y}Y}\partial_z Y = 
 -\frac{1}{2}\frac{\partial\bar{\cal W}}{\partial\bar Y}
 \label{five}\ ,
\ee 
which in the present case reduces to 
\be 
 \partial_{\hat z} Y = - \bar Y + \bar Y^{-3}\ , \label{jBPS}
\ee
on introducing a dimensionless coordinate $\hat z = |m_1| z$.       

For configurations satisfying the Bogomol'nyi equation, the junction 
tension $T_j$ saturates the BPS bound for the (1/2,1/2) central
charge \cite{dwj2},
\be
 T_j = -\frac 1 2 \sqrt{\La_1^5m_1^{-1}}\,\oint a_k dx_k\,, \;\;\;\;\;\;\;\;\;\; 
 a_k = i\bar Y\stackrel{\leftrightarrow}{\partial_k} Y\,,\qquad k = 1,2\ ,
\label{seven}
\ee
where the integral runs over a large contour in the
$xy$ plane. In the problem at hand it is convenient to choose a
rectangular contour which must lie in the plane of  Fig.~3.
%\ref{ff4}. 
Then, on the vertical sides 
of the contour (i.e. those parallel to the wall)
the field $Y$ is essentially constant; therefore, $a_k=0$.
Moreover, on the horizontal sides of the contour (i.e. those perpendicular to the wall)
only the phase of
$Y$ changes, and so 
\be
 a_k = - \,\partial_k \tilde\theta\,,  
\label{nine}
\ee
where the phase $\tilde\theta$ was defined in Eq. (\ref{wtraj}).
We do not need to know precisely how $\tilde\theta$ depends on $x$ since 
the contour 
integral can be done directly,
\be
  \,\oint a_k dx_k = - \Delta \tilde\theta \,,\;\;\;\;\; {\rm with}\;\; 
    \Delta \tilde\theta = 2\pi\,,
\label{ten}
\ee
where the numerical result holds 
for the field configuration depicted in Fig.~3. Thus,
\be
T_j = \pi \, \sqrt{\left|\frac{\Lambda_1^5}{m_1}\right|}\ .
\label{eleven}
\ee
This tension is determined\,\cite{dwj2} by the (real) central charge
in the anticommutator $\{\bar Q, Q\}$ as in (\ref{sc}). However, in contrast to the wall 
tension, it is not holomorphic in parameters, and thus we cannot 
extrapolate this result to large $m$ where one recovers pure SYM. 

Note that we can interpret (\ref{eleven}) as implying that the
thickness of the junction in the $y$ direction is of the same order
as the thickness of the wall itself in the $x$ direction and is
large, $\sim m^{-1}$. Because of this fact the presence of an 
unknown core in the wall junction (which is inevitable since the 
$Y$ trajectory runs through the strong coupling domain)
is unimportant numerically since the relevant scale is $\La_1$. 
However, this point necessarily means that the junction cannot
fully be described within this effective theory, and we can ask
whether the worldvolume perspective may help in this regard.

\subsection{Resolving the singularity on the worldvolume}

The preceding analysis indicates that the $N_f=1$ system is
rather inadequate for describing the detailed structure of the
junction solution. In particular, the boundary conditions at infinity
in the plane transverse to the axis of symmetry ensure the
following symmetry of the solution,
\be
  Y(z) \rightarrow \bar Y(\bar z)\ .
\ee
Along with continuity, this implies that the field $Y^2$ must 
pass through zero at the core of the junction. Since the 
potential diverges at this point, we see that the description 
breaks down within the shaded domain sketched at the right in Fig.~3, and 
we cannot expect to find a solution (even numerically) in this region.
A similar singularity is seen to arise, for similar reasons, for
analogous 2-wall junctions for higher $N$.

It is interesting then to address this question directly from the
worldvolume point of view, by adding an additional light flavor so
that the 2 component walls arise from the dynamics of a $\cp^1$
sigma model, or more precisely a massive sigma model where the
mass term is identified with $|\pi\De m|/2$ as discussed in 
the previous section. This theory possesses 1/2-BPS kink solitons, and
it is natural to identify these kinks as the worldvolume description of 
1/4-BPS 2-wall junctions. We will now provide evidence for this identification
by verifying that the kink tension reproduces the tension of the junction,
given in (\ref{eleven}), in the appropriate limit.

Using complex coordinates for $S^2$, as introduced in
(\ref{compcoord}), the bosonic sector of the massive $\cp^1$ sigma model becomes 
\be 
 {\cal L} = \frac{2R_{\widetilde{\cal M}}}{(1+|w|^2)^2} \Big[|\ptl_{\mu} w|^2 
+ \frac{1}{4}\, |\pi \De m|^2 |w|^2\Big],
\ee
where $R_{\widetilde{\cal M}}$ is the K\"ahler parameter denoting the size of the 
reduced moduli space. When $|\De m|$ is large relative to any dynamically
generated scale, the theory has classical vacua at $w=0,\infty$. 
Using the coordinate relation from (\ref{compcoord}), $w = \tan\frac{\xi_0}{2}e^{i\ph_0}$,
one finds that classical BPS kink solutions exist which 
satisfy (yet) another sine-Gordon equation \cite{dorey98}
\be
 \ptl_y \xi_0 = \pm \frac{1}{2}\pi |\De m| \sin\xi_0\,,  \;\;\;\;\;\;\;\;
   \ptl_y \ph_0 = 0\ . \label{juncSG}
\ee
The corresponding tension of the junction
is given by 
\be
 T_j = \frac{1}{2}\pi |\De m | (2R_{\widetilde{\cal M}}) + {\cal O}({\La_{\rm wv}})\ , \label{kinktens}
\ee
where for the moment we assume $\De m$ is large and so provides
the dominant mass scale.

In the present case $R_{\widetilde{\cal M}}$ was computed in
(\ref{Mscale}), and we obtain
\be
 T_j = \pi \left|\frac{|m_1| - |m_2|}{\sqrt{|m_1m_2|}}\right| \La_2^2\ . \label{Tj2}
\ee
A simple check on this result follows on integrating out
one of the flavors. On sending $m_2 \rightarrow \infty$,
we must keep $\La_1^5 = m_2 \La_2^4$ fixed, so that
\be
  T_j =  \pi \left|\frac{|m_1| - |m_2|}{\sqrt{|m_1m_2|}}\right| 
   \sqrt{\frac{\La_1^5}{|m_2|}}
     \; \stackrel{\scriptscriptstyle 
m_2\rightarrow\infty}{\displaystyle \longrightarrow} \;
    \pi \sqrt{\frac{\La_1^5}{|m_1|}}\ , \label{Tjlim}
\ee
which agrees precisely with the result obtained earlier in Eq.~(\ref{eleven})
from a direct analysis of the 1-flavor model, despite being derived in the small $\De m$ regime. 
It is worthy of note that the earlier determination
that the worldvolume `real' mass perturbation was indeed a 
real parameter also finds a nice consistency check in this expression. The resulting junction
tension depends non-holomorphically on $m_1$ as one expects from the bulk
point of view.

Examining (\ref{Tj2}) we observe that as the mass splitting is reduced
we become sensitive to quantum effects on the worldvolume, and
indeed this is to be expected as the bulk theory is also strongly
coupled for $m_1 \sim m_2$. In this regime, the junction configuration
is still described by a $\cp^1$ kink, and we conclude that
the solution will be nonsingular whenever the worldvolume IR dynamics is sufficient to generate
a mass gap. For example, it is sufficient to compactify one of the 
spatial dimensions on an $S^1$ of circumference $L$. The effective 1+1D
dynamics then generates a dynamical scale of the form
\be
 \La_{\rm wv} = \mu \exp\Big(\! - \frac{2\pi}{g^2(\mu)} \Big)\,, \;\;\;\;\; 
  g^2(\mu) = \frac{1}{R_{\widetilde{\cal M}}L}\ , \label{LaW}
\ee
and it is this parameter which enters the (1/2,1/2) central
charge and sets the tension, or `effective mass', of the junction, 
reduced now to a localized soliton. This determines a 
contribution to $T_j$ which is necessarily independent of the 
contribution from the walls. Unfortunately, since the relevance of 
such a  worldvolume scale only becomes apparent on entering the strong coupling
regime when $m_1 \sim m_2$ it is difficult to make any concrete identification
with bulk 3+1D parameters.\footnote{In contrast, an identification of the quantum scale 
is possible within the analogous $\cp^{N-1}$ worldsheet dynamics of `nonabelian' vortices 
in the \ntwo Higgs phase \cite{vortex}.} Nonetheless, this scale does
have a direct physical interpretation in 3+1D as the intrinsic junction tension.

The enhancement of worldvolume supersymmetry for the reduced wall
moduli space also has important consequences for this identification of 1/4-BPS bulk junctions
with 1/2-BPS worldvolume kinks. In particular, while the junction preserves only one of the
bulk supercharges, the kink preserves in addition one of the supernumerary charges
present on the  worldvolume. Moreover, since the junction not 
only preserves two worldvolume supercharges, but also breaks two, it necessarily exhibits 
two fermionic zero modes. Recalling the discussion at the start of Sect.~IV, we see that 
this is not the minimal `chiral' content that one would anticipate based on the breaking of 
bulk supersymmetry. In actual fact, the 
kink solutions also have two bosonic moduli, the center of 
mass position and the phase $\ph=\ph_0$ as is apparent from (\ref{juncSG}). Thus
the moduli space is two-dimensional
\be
 {\cal M}_{\rm kink} = \R \ \times \ \widetilde{\cal M}_1\ ,
\ee
where the reduced moduli space is $\widetilde{\cal M}_1 = S^1$. Note that only
one of these bosonic moduli
-- the translational mode -- would have been anticipated from a consideration of 
the bulk kinematics. We see that the bosonic and fermionic moduli form two (0,1) chiral multiplets
and two bosonic singlets. On the worldvolume, this structure is enforced by the broken
supersymmetry. However, from the bulk point of view the second (0,1) chiral multiplet
and singlet are not required by supersymmetry considerations, but presumably correspond to a Goldstone
multiplet arising through the breaking of flavor symmetry, as is the case for the 
wall itself.   

In this context, the \ntwo worldvolume SUSY
resolves an apparent paradox that arises when one
tries to verify that these junction solutions are BPS saturated at the
quantum level. In particular, were the worldvolume to possess only
\none SUSY, putative BPS junctions 
would have to be realized as one-component multiplets. 
An index to count such multiplets was introduced in \cite{lsv}, 
which is formally expressed as 
\be
 \nu_{\rm LSV} = \frac{1}{2{\cal Z}}\, \{{\rm Tr}\,Q_{\rm broken}\}^2\ ,
\ee
in terms of the broken supercharge. An observation of \cite{lsv} which is 
particularly relevant here is that the index necessarily vanishes when the fermion parity
$(-1)^F$ is well-defined. This indicates that the multiplet
is generally reducible (containing two states)
and can lift from the BPS bound. In the present case, as noted above, the 
kinks have an even number of bosonic, and consequently fermionic, moduli. On quantization, the latter furnish 
a representation of the Clifford algebra, in this case $\ga^i=\si_1,\si_2$, from which we can
construct $\ga_5 = \si_3$ which represents $(-1)^F$.
Thus one would necessarily conclude that no short \none multiplets
are allowed and there would be no reason to expect that these junctions
should saturate the 1/4-BPS bound in 3+1D. This would be rather puzzling, and
indeed as we have discussed this problem is resolved due to the 
enhanced SUSY on the reduced moduli space so that the junctions
lie in BPS multiplets of \ntwo SUSY, and are instead counted by the
CFIV index on the worldvolume.

\subsection{Extensions for SU($N$)}

An immediate technical advantage of the realization of junctions as
BPS kinks on the wall worldvolume, is that we can utilize our knowledge 
of these configurations for arbitrary $N$ to infer analogous results for
junctions, which are in fact rather difficult to obtain directly.
Thus we now identify 1/4-BPS 2-wall junctions in the SU($N$) theory
with $N_f=N$ with kinks in the worldvolume $\cp^{N-1}$ sigma model
deformed by the relevant real mass terms. We will limit our remarks here to
two issues, namely the multiplicity of 2-wall junctions, and their tension.

In order to make this discussion concrete we must again resort to compactifying the
theory on a circle to ensure that the low energy effective theory on the wall worldvolume 
is 1+1-dimensional and develops a mass gap.
We can then vary the quark masses across the range where the dynamical scale
$\La_{\rm wv}$ becomes important and, for example, sit in the strong coupling 
region where $\De m \ll \La_{\rm wv}$. Note that whether or not this restriction changes 
the physical conclusion is tied to the question of whether entering the strong coupling
domain in the bulk effectively induces a mass gap within the 
(decompactified) worldvolume theory.

Due to \ntwo SUSY, the junction multiplicity is formally given by the CFIV index
as noted above.  Focusing just on minimal walls for arbitrary $N$, a generic 
intersection between two of the $N$ possible walls, will connect walls differing 
by $p$ units of phase -- we will refer to this as a $p$-junction. The number of
$p$-junctions is formally (on compactification on $S^1$, and taking the
limit $|\De m|\rightarrow 0$)
\be
 \nu_{\rm CFIV} = \left( \begin{array}{c} N \\ p \end{array}\right).
\ee
In practice, as noted above, the result can be
somewhat different in the limit  $|\De m|\gg \La_{\rm wv}$, which may
in fact be the only accessible regime in 2+1D. In particular, 
in this regime in 1+1D one can turn on an arbitrary integer `dyonic' charge \cite{dorey98},
due to a coupling to the corresponding U(1) current in the superalgebra \cite{ls},  
although this is also reflected in a change in the mass. Only a certain number
of these states survive (as above) in the limit $|\De m|\rightarrow 0$,
due to the presence of marginal stability curves \cite{dorey98}. 

Turning to the tension, in the hierarchical regime $|\De m|\gg \La_{\rm wv}$, the
result is a natural generalization of (\ref{kinktens}) determined by the
various real mass terms. More interesting perhaps is that, 
within the compactified regime with $|\De m|\ll \La_{\rm wv}$,
the result translated directly from that for $\cp^{N-1}$ kinks is 
\be
 T_{p} = \La_{\rm wv} \sin \frac{2\pi p}{N}\ .
 \label{dsf}
\ee
Although this is a rather familiar formula in the context of wall-like solitons,
it takes on an interesting new interpretation here as the junction is a string-like
source, albeit wrapped on a small circle in the present construction. Since
this result is naively protected by the enhanced \ntwo worldvolume 
SUSY,\footnote{This statement requires some caution as reference to Eq.~(\ref{Tjlim}) indicates that,
due to the embedding, the dependence of the tension on the mass scale in the hierarchical limit is
not holomorphic. This is in accord with expectations for the junction charge in the bulk.}
we see that the wall junctions actually realize the `sine formula' for the ratio of 
`string' tensions for differing values of $p$ first observed for strings in 
softly broken \ntwo SYM by Douglas and Shenker \cite{ds}.\footnote{A heuristic model relating 
the wall tension and the string tension (\ref{dsf}) was discussed recently in \cite{aams}.}

This structure is not expected to apply to generic $p$-strings in confining vacua of
\none SYM, since these states are non-BPS, but here we find a situation where the
sine-formula appears to be exact, due to the enhanced SUSY on the worldvolume.
However, we should reiterate that this discussion has been framed within a specific scenario.
If we decompactify the extra spatial dimension, then to retain control over the vacuum
structure, one needs to reintroduce a hierarchy for the quark masses. The kink spectrum, and also the
tension, then
changes considerably on moving outside a `curve of marginal stability', and 
many more states are present classified by U(1) charges associated with the residual
abelian flavor symmetries in the hierarchical case. After decompactification, if we try to
remove the hierarchy the system re-enters a strong coupling regime that at present
appears intractable.

\section{Concluding Remarks}

In this paper we have presented a detailed exploration of the worldvolume 
moduli space dynamics of 1/2-BPS domain walls in \none SQCD with gauge group
SU(2) and $N_f=2$ flavors. We have also discussed how novel 1/4-BPS 2-wall
junctions may be realized as kinks within the worldvolume theory. We concentrated 
on the SU(2) example where much of the analysis could be performed explicitly, but we anticipate
that most of the conclusions should extend to the generic SU($N$) case
with $N_f=N$ flavors. In particular, the appearance of an enhanced \ntwo worldvolume
supersymmetry on the reduced moduli space is essentially guaranteed by the
corresponding construction as a K\"ahler quotient. In this concluding
section, we will make a couple of more speculative remarks on
localized worldvolume solitons which may (or may not) find a bulk interpretation.

The \ntwo algebra in 2+1D includes, in addition to a tensorial central charge
for the 1+1D kink which we have interpreted as a 2-wall junction, a Poincar\'e
invariant charge supported by localized lump solitons. When the 
worldvolume theory is naturally embedded in the relevant linear sigma model, lumps 
are realized as semi-local vortices. Consequently, with reference to the
interpretation of such walls as D-branes for SYM strings \cite{witten97}, 
it is tantalizing to speculate that these configurations may have a relation 
to the endpoints of SQCD strings.\footnote{Note that worldvolume 
vortices on BPS walls were shown to represent string endpoints 
in gauge theories with eight supercharges \cite{msay} (see also \cite{HKsigma}).}
Note in particular that at energy scales 
below the UV cutoff on the wall, of order $1/\mu$, such strings are stable 
to quark pair production. The result $\nu_k$ for the wall multiplicity is
also consistent with the interpretation that 1-walls lie in the fundamental 
representation of SU(N), or more generally admit an action
of the corresponding Weyl group, and form antisymmetric bound states.
Moreover, although these configurations are BPS on the worldvolume, they would indeed 
be non-BPS within the bulk. 

In spite of these intriguing hints it seems difficult,
for several reasons, to make a precise 
identification of this type. For example, lumps carry
integer charges $\pi_2(\C P^1) = \Z$, rather than charges under $\Z_N$
that one might associate with the center of the gauge group. A contraction 
of the charge lattice, $\Z \rightarrow \Z_N$, might occur due to physics occurring
above the worldvolume UV cutoff, but there is another more significant roadblock in the way of 
a quantitative
study of this question. This is the fact that in the hierarchical mass regime where the 
theory is tractable there is a potential on the moduli space. In the presence of
such a deformation, lumps are no longer stable, via Derrick's theorem, unless 
one turns on additional U(1) charges. Such time-dependent $Q$-lumps are known, for $N>2$, to
have a ring-like structure and locally carry the junction charge \cite{at2}. Thus these
configurations appear as domain wall bubbles on the worldvolume. It is far from
clear what may happen to these configurations as one sends $\De m \rightarrow 0$ 
and returns to strong coupling, and this hinders a direct bulk interpretation.

In a similar regard, we can also speculate about configurations which one
might dub `junctions of junctions'. In particular, the counting argument
for junctions described above suggests that, even in the minimal SU(2)
case, there are two inequivalent junctions. One may then anticipate
that a further intersection, now within the worldvolume, 
would be possible -- a 1/4-BPS state on the
worldvolume (since it sources both the kink and lump central charges), 
but again non-BPS in the bulk.

\begin{acknowledgments}
We would like to thank J. Gauntlett and A. Yung for helpful discussions and comments on the manuscript.
A.R. thanks the FTPI at the University of Minnesota for their generous hospitality while part of this
work was completed.
The work of M.S. and A.V. was supported in part by DOE grant DE-FG02-94ER408.
\end{acknowledgments}

\appendix

\section{On central charges in \boldmath${D=2,3}$ and 4}

In this Appendix we will briefly discuss the central charge (CC) content of 
the superalgebras in $D=2,3$ and $4$ relevant to this paper, and their inter-relations. 

One may recall that some time after the minimal four dimensional superalgebra was first written 
down by Golfand and Likhtman \cite{A1}, central charges were introduced algebraically
as Poincar\'e invariant, and thus scalar, elements of the superalgebra commuting
with all the other generators \cite{A2,A3} (see also \cite{Grisaru}). Their dynamical role was 
subsequently made apparent by Witten and Olive \cite{A4}, who showed that such charges 
are supported by the topological charges of solitons. While it was appreciated for some time
that not all central charges are Lorentz scalars (see e.g. \cite{vHvP}), the dynamical role
of these additional tensorial charges was not fully understood
until somewhat later, when they were shown to be nonzero in the presence of
extended objects ($p$-branes) within supergravity \cite{agit} (see also \cite{at}). 
Their occurrence in \none SYM in $D=4$ (via a quantum anomaly) was first observed 
in \cite{DSone}.

We will concentrate on the algebras in $D=2,3$ and 4, for which the analyses in 
\cite{gates} for $D=2$, and \cite{A7} (see also \cite{dwj2})
for $D=3$ and 4, are particularly relevant. While most of what follows
comprises review material collected here for completeness, we will extend the 
discussion in \cite{A7} of vectorial central charges, namely those with the 
Lorentz structure of $P_\mu$.

\subsection{Minimal SUSY}

Limiting ourselves to two, three and four dimensions we observe that
the minimal number of supercharges is 2, 2 and 4,
respectively. Two-dimensional theories with a single
supercharge, although algebraically possible,
require the loss of $F$ and $(-1)^F$. Therefore, if one wishes
to keep the distinction between `bosons' and `fermions',
the minimal number of supercharges in $D=2$ is two. 

Working in a real representation with $\nu_Q$ supercharges,
it is clear that, generally speaking, the maximal possible number of CC's is
determined by the dimension of the symmetric matrix $\{Q_i, Q_j\}$ of 
size $\nu_Q\times \nu_Q$, namely,
\beq
\nu_{\rm CC} = \frac{\nu_Q (\nu_Q +1)}{2}\ .
\eeq
In fact, $D$ anticommutators have the Lorentz structure of the
energy-momentum operator $P_\mu$. Therefore, up to $D$ central charges
could be absorbed in $P_\mu$. However, in particular situations
this number can be smaller, since although algebraically
the corresponding CC's have the same structure as $P_\mu$, they are dynamically 
distinguishable. The point is that $P_\mu$ is uniquely
defined through the conserved and symmetric energy-momentum 
tensor of the theory. 

The total set of CC's can be arranged by classification with respect to their 
Lorentz structure. Below we will present this classification for minimal 
supersymmetry in $D=2,3$ and 4. We then consider the extended \ntwo supersymmetry algebras in
$D=2$ and $D=3$ obtained via dimensional reduction from $D=4$, and consider the
analogous decomposition in terms of Lorentz and $R$-symmetry representations.

\bigskip
\noindent$\bullet\;D=2$
\bigskip

Consider two-dimensional theories with two supercharges.
From the discussion above, on purely algebraic grounds,
three CC's are possible: 
\be
\{Q_\al,Q_\beta\} = 2(\ga^\mu\ga^0)_{\al\beta}(P_\mu+Z_\mu)+
 2i (\ga^5)_{\al\beta}Z\ ,
\ee
one Lorentz-scalar $Z$ and a two-component vector $Z_{\mu}$. The latter case 
would require the existence of a vector  order parameter taking distinct values in different vacua. 
This will break  Lorentz invariance and supersymmetry of the vacuum state.
Limiting ourselves to supersymmetric vacua we conclude that only one
(real) Lorentz-scalar central charge is possible. This central 
charge is relevant to kinks in \none theories.

\bigskip
\noindent$\bullet\;D=3$
\bigskip

The central charge allowed in this case is a Lorentz-vector $Z_{\mu}$,  
i.e.
\be
 \{Q_\al,Q_\beta\} = 2(\ga^\mu\ga^0)_{\al\beta}(P_\mu+Z_\mu)\ ,
\ee
which we should arrange to be orthogonal to $P_\mu$. By an appropriate choice of reference frame
it can always be cast in the form $(0,0,1)$. In fact, this is
the central charge of the previous section elevated by one dimension.
It is associated with a domain wall (or string) oriented along the
second axis.
 
\bigskip
\noindent$\bullet\;D=4$
\bigskip

Maximally one can have 10 CC's which are decomposed
into Lorentz representations as (0,1) + (1,0) + (1/2, 1/2):
\bea
 \{Q_{\al},\bar{Q}_{\dot \al}\} &=& 2(\ga^\mu)_{\al\dot\al}(P_\mu+Z_\mu)\,, \nonumber\\[1mm]
 \{Q_{\al},Q_{\beta}\} &=& (\Si^{\mu\nu})_{\al\beta}\bar{Z}_{[\mu\nu]}\,,\nonumber \\[1mm]
 \{\bar Q_{\dot\al},\bar Q_{\dot \beta}\} &=& (\bar \Si^{\mu\nu})_{\dot \al \dot \beta} Z_{[\mu\nu]}\,,
\eea
where $(\Si^{\mu\nu})_{\al\beta} = (\si^\mu)_{\al\dot\al}(\bar{\si}^\nu)^{\dot\al}_\beta$ is a chiral
version of $\si^{\mu\nu}$ (see e.g. \cite{SVinst}).
The antisymmetric tensors $Z_{[\mu\nu]}$ and $\bar{Z}_{[\mu\nu]}$ are 
associated with domain walls, and reduce to a complex number and a 
spatial vector orthogonal to the domain wall.
The (1/2, 1/2) CC $Z_\mu$ is a Lorentz vector orthogonal to
$P_\mu$. It is associated with strings (flux tubes), and
reduces to one real number and a three-dimensional unit
spatial vector parallel to the string.

\subsection{Extended SUSY}

We will limit our attention here to exploring the reduction of the minimal SUSY algebra
in $D=4$ to $D=2$ and 3, namely the \ntwo SUSY algebra in those dimensions. 
As should be clear from the discussion above, the maximal number of CC's is of course
the same, and the only distinction we must make is to provide a decomposition into both
Lorentz and $R$-symmetry irreps.

\bigskip
\noindent$\bullet$ \ntwo in $D=3$
\bigskip

The superalgebra can be decomposed into Lorentz and $R$-symmetry tensorial structures as follows:
\be
 \{Q_\al^i,Q_\beta^j\} = 2(\ga^\mu\ga^0)_{\al\beta}[(P_\mu+Z_\mu)\de^{ij} + Z_{\mu}^{(ij)}]
 +2(\ga^0)_{\al\beta}Z^{[ij]}\ , \label{tdecomp}
\ee
where $\ga^0$ is the charge conjugation matrix. The maximal set of 10 CC's
enter as a triplet of spacetime vectors $Z_{\mu}^{ij}$ -- which we decompose into an 
$R$-symmetry singlet trace
term, denoted $Z_\mu$,  
and a trace-free symmetric combination $Z_\mu^{(ij)}$ -- 
and a singlet $Z^{[ij]}$. The singlet CC is associated with vortices (or lumps),
and corresponds to the reduction of the (1/2,1/2) charge or the $4^{th}$ component of the 
momentum vector in $D=4$. The $R$-symmetry singlet $Z_\mu$ is algebraically 
indistinguishable from the momentum and is equivalent to 
the vectorial charge in the \none algebra. The traceless symmetric combination 
$Z_\mu^{(ij)}$ can be reduced to a complex number and vectors specifying the orientation
of a co-dimension one source.  We see that these are the direct 
reduction of the (0,1) and (1,0) wall charges in $D=4$.

\bigskip
\noindent$\bullet$ \ntwo in $D=2$
\bigskip

Lorentz invariance now provides a much weaker constraint, and one can in principle
consider different $(p,q)$ superalgebras with $p\neq q$. We will focus here only
on the nonchiral ${\cal N}=(2,2)$ case corresponding to dimensional reduction of the
\none $D=4$ algebra. The tensorial decomposition is as in (\ref{tdecomp}), but
with the decomposition of $D=3$ spacetime vectors into $D=2$ vectors and a 
singlet,
\be
\{Q_\al^i,Q_\beta^j\} = 2(\ga^\mu\ga^0)_{\al\beta}[(P_\mu+Z_\mu)\de^{ij} + Z_{\mu}^{(ij)}]+
 2i (\ga^5\ga^0)_{\al\beta}(\de^{ij}Z + Z^{(ij)})
 +2(\ga^0)_{\al\beta}Z^{[ij]}\ ,
\ee
We discard all vectorial charges $Z_{\mu}^{ij}$ in this case for the same reasons 
as noted above in the \none case, namely they would imply SUSY breaking in the vacuum. 
This leaves two singlets $Z^{(ij)}$, which are the reduction of the domain wall charges in $D=4$
and correspond to topological kink charges, and two further singlets $Z$ and $Z^{[ij]}$, arising
via reduction from $P_2$ and the vortex charge in $D=3$. The latter charges also arise for
kinks in the presence of twisted mass terms \cite{gates}.

\end{document}